\documentclass[useAMS,usenatbib]{mn2e}
\usepackage{amsmath,amssymb,graphicx,color,longtable}
\newcommand{\kg}{$\kappa,\gamma$~}

\title[Microlensing flux ratio predictions for Euclid]{Microlensing flux ratio predictions for Euclid}

\author[G. Vernardos]{G. Vernardos$^{1}$\thanks{E-mail: gvernard@astro.rug.nl}\\
	$^{1}$Kapteyn Astronomical Institute, University of Groningen, PO Box 800, NL-9700AV Groningen, the Netherlands\\
	}
\date{Accepted XXX. Received YYY; in original form ZZZ}
\pubyear{2017}

\begin{document}
\label{firstpage}
\pagerange{\pageref{firstpage}--\pageref{lastpage}}
\maketitle

\begin{abstract}
Quasar microlensing flux ratios are used to unveil properties of the lenses in large collections of lensed quasars, like the ones expected to be produced by the Euclid survey.
This is achieved by using the direct survey products, without any (expensive) follow-up observations or monitoring.
First, the theoretical flux ratio distribution of samples of hundreds of mock quasar lenses is calculated for different Initial Mass Functions (IMFs) and Sersic radial profiles for the lens compact matter distribution.
Then, mock observations are created and compared to the models to recover the underlying one.
The most important factor for determining the flux ratio properties of such samples is the value of the smooth matter fraction at the location of the multiple images.
Doubly lensed CASTLES-like quasars are the most promising systems to constrain the IMF and the mass components for a sample of lenses.
\end{abstract}


\section{Introduction}
\label{sec:intro}

Quasar microlensing is a phenomenon that can carry physical information from very small angular scales [$\mathcal{O}(10^{-6})$ arcsec], hardly accessible in other ways.
It allows one to investigate from the structure of the lensed source down to the size of the central accretion disc, to properties of the lensing galaxy like the local amount of the smooth and compact matter components.
Determining the partition of the lens mass in compact (stellar) and smooth (dark) matter components can shed light into dark matter halo creation and the accompanying star formation.
Measuring properties of the accretion disc can help to constrain theories of supermassive black hole growth and evolution.
Finally, cosmological models can be tested by measuring time delays in lensed quasars and supernovae (in this case microlensing introduces noise to the sought after measurement, rather than being the signal itself).

The study of lens galaxy mass distribution and the partition between stellar and dark matter is an active field of research.
Currently, there is evidence that the total lens potential is very close to the Singular Isothermal Ellipse model \citep[SIE,][]{Kassiola1993,Rusin2003,Koopmans2009,Tortora2014,Oldham2018}, at least for a lens constituted by a giant elliptical galaxy.
To disentangle the stellar matter contribution from the total, one may measure the light profile of the lens and derive an estimate for the baryonic matter content based on a mass-to-light ratio \citep{Grillo2009,Auger2009,Courteau2014}.
However, constraining the two matter components in this way is strongly affected by the fundamental Initial Mass Function (IMF) - dark matter fraction degeneracy \citep[e.g. see][]{Oguri2014}, which, in short, states that the same lens potential and light distribution can be produced by a low dark matter fraction and many low-mass (and less luminous) stars, or a high dark matter fraction and fewer but brighter stars.
The partition of the lens mass will eventually determine the properties of any resulting microlensing signal (in addition to the size of the source that is being lensed).

One manifestation of microlensing is its effect on flux ratios between the multiple images of a lensed quasar.
These ratios are fixed by the magnification of each image produced by the total lensing potential (the macromodel).
Such observations can be used to study the structure of the lens potentials \citep[][]{Mutka2011} or perform cosmological measurements \citep{Mortsell2006}.
However, the small scale perturbations (graininess) of the total lensing potential due to stellar-mass objects cause the magnification of each image to deviate independently.
These wavelength dependent deviations \citep[stronger for shorter wavelengths,][]{Rauch1991,Agol1999} can be used to unveil the structure of the background quasar \citep[e.g.][]{Bate2008,Poindexter2008,Mediavilla2011a,Sluse2011}.
Alternatively, substructure ($>10^6$ M$_{\odot}$) in the lens can cause the flux ratios to deviate from their macromodel values as well \citep{Mao1998,Dalal2002,Metcalf2002}, albeit with different characteristics compared to microlensing \citep{Inoue2016}.
For example, \cite{Vernardos2014a} found that microlensing produces demagnification along the critical line (see equation \ref{eq:mag} and caption of Fig. \ref{fig:pspace}), extending the result of \cite{Schechter2002}.
 
So far, there have been $\sim 500$ lensed quasars discovered \citep[see][for an up-to-date list]{Ducourant2018}, mostly serendipitously in surveys.
For subsequent microlensing studies, detailed follow-up has been performed with customized observations for a few tens of targeted systems \citep[e.g.][]{Jimenez2015b}.
Future all-sky surveys carried out by instruments such as Euclid \citep{Laureijs2011} or the Large Synoptic Survey Telescope \citep[LSST,][]{LSST2009} are expected to discover thousands of such objects (see \citet{SLSWG2018} and \citet{Oguri2010}, hereafter OM10), and provide homogeneous sets of imaging and monitoring data.
However, due to the long timescales of microlensing variability \citep[see][]{Mosquera2011b} and the possible presence of broadband contaminants \citep[e.g. see][]{Bate2018} it is likely that expensive follow-up observations will still be required to get the most out of each system.
Additionally, resolving any presence of lens substructure, which can mimick the microlensing effect, will require extensive modelling (e.g. see \citealt{Vegetti2009a} for modelling of extended lensed images and \citealt{Xu2015} for point like sources).

However, a sample of a few hundred objects will wash out the effect of any outliers that may be showing extreme microlensing behaviour, be affected by the presence of substructures, or both.
It will, therefore, allow one to consistently and systematically study the effect of microlensing throughout the relevant parameter space (see next section).
Although the microlensing simulations required are very computationally demanding, the publicly available GERLUMPH\footnote{{\tt http://gerlumph.swin.edu.au}} parameter survey already provides exactly such a suite of simulations, unique in its kind \citep{Vernardos2014a,Vernardos2014b,Vernardos2015}.

What are the theoretically expected microlensing properties of a large sample of lensed quasars?
What can we learn from the Euclid data alone without the need for any expensive follow-up observations?
This work attempts to answer these questions.
In Section \ref{sec:method} the various model components and techniques are presented in order to get a microlensing flux ratio probability distribution.
Such distributions are then analyzed in Section \ref{sec:results}, in light of the lensing galaxy properties.
Finally, discussion and conclusions are presented in Section \ref{sec:discuss}.

Throughout this paper a fiducial cosmological model with $\Omega_{\rm m}=0.26$, $\Omega_{\rm \Lambda}=0.74$, and $H_0=72$ km s$^{-1}$ Mpc$^{-1}$ is used.

\section{Sample and method}
\label{sec:method}
The statistical properties of microlensing flux ratios are examined over collections of different simulated samples, each consisting of $\sim2300$ multiply imaged quasars, for which mock observations are produced.
The different observational signatures of such ensembles are set by:
\begin{enumerate}
	\item the total mass distribution of the lens,
	\item its stellar mass distribution,
	\item the source, i.e. the quasar accretion disc,
	\item and the observing instrument.
\end{enumerate}
The models employed to describe each of these characteristics for different simulated lensed quasar populations are described in the following sections.
The mock observations are generated using the specifications of the upcoming Euclid space telescope.
The final products of the method presented here are various microlensing flux ratio probability distributions (analyzed in Section \ref{sec:results}), which can be used to directly probe the partition of matter in smooth and compact components.

\subsection{A sample of mock lenses}
\label{sec:mocks}
The mass catalog of OM10 and the accompanying tools\footnote{{\tt https://github.com/drphilmarshall/OM10}} are used to create a sample of strongly lensed quasars.
The lenses are assumed to be elliptical galaxies having lensing potentials described by the SIE family of mass models with an additional external shear (also referred to as the macromodel).
The survey specifications for the OM10 catalog are set to match those of the Euclid VIS instrument in the wide survey mode, having: a limiting magnitude of 24.5, an area of 15,000 deg$^2$, and a resolution better than 0.2 arcsec.
This results in 2346 lensed quasars, 321 (14 per cent) of which are quadruply imaged \citep[see also][]{SLSWG2018}.
The source and lens redshifts of the sample are shown in Fig. \ref{fig:z_hist}.

\begin{figure}
	\includegraphics[width=0.5\textwidth]{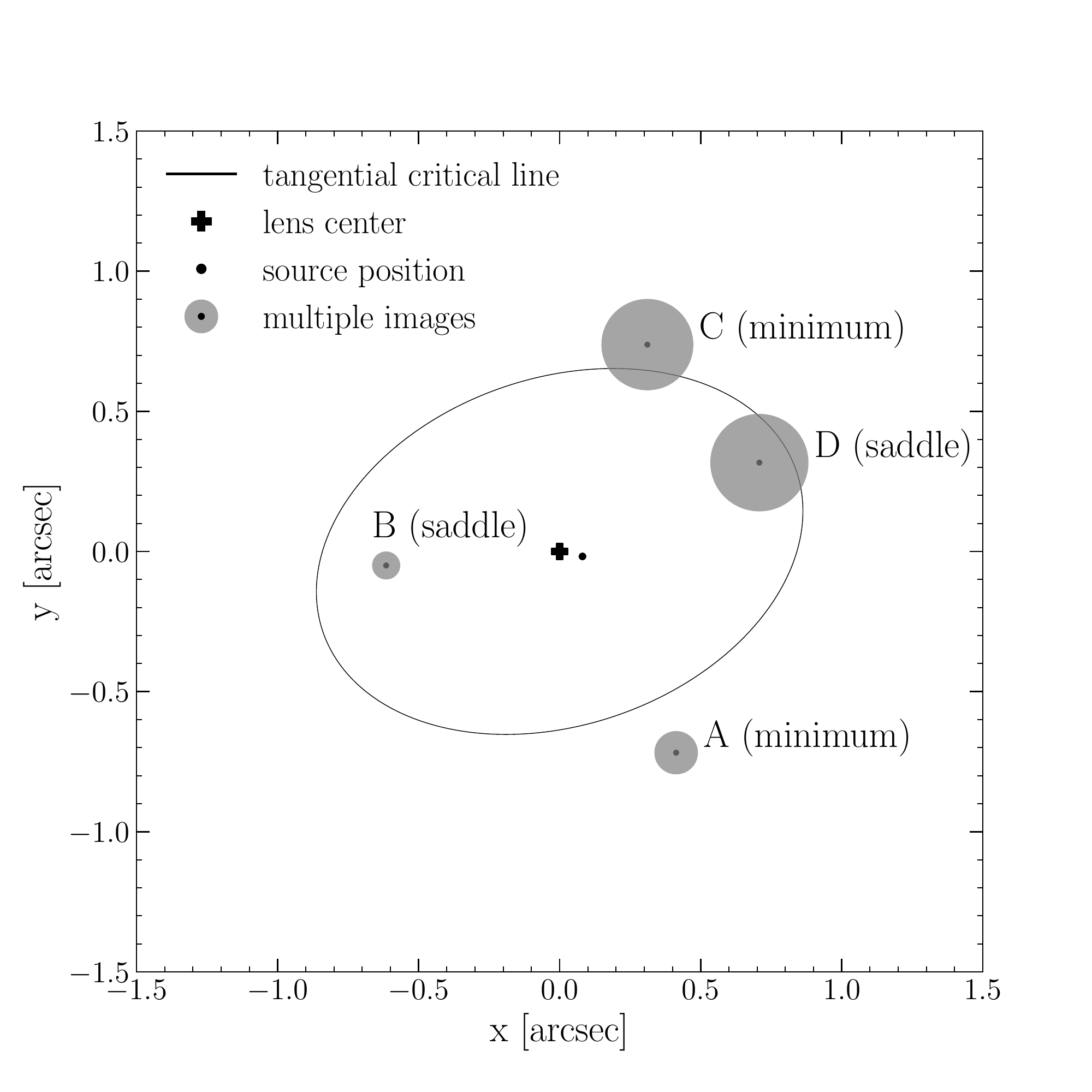}
	\caption{Image configuration of a fiducial quadruply imaged quasar. The parameters of the system are given in Table \ref{tab:example}. The multiple images are shown as circles with their radius proportional to their magnification relative to image B (the least magnified one). Images C and D form a close pair and are highly magnified (see Table \ref{tab:example_images}). The tangential critical line of the lens, which roughly corresponds to $\theta_{\rm Ein}$, is also shown.}
	\label{fig:example}
\end{figure}

An example of a quad image configuration is shown in Fig. \ref{fig:example}, with the system parameters listed in Table \ref{tab:example}.
The Einstein radius is given by:
\begin{equation}
\label{eq:theta_rein}
\theta_{\rm Ein} = 4 \pi \bigg( \frac{\upsilon_{\rm disp}}{c} \bigg)^2 \frac{D_{\rm LS}}{D_{\rm S}} \, \lambda(q),
\end{equation}
where $\upsilon_{\rm disp}$ is the velocity dispersion of the lens, $c$ is the speed of light, and $D_{\rm LS},D_{\rm S}$ are the angular diameter distances from the lens to the source and from the observer to the source respectively.
To account for the three-dimensional non-spherical shapes of lensing galaxies (the SIE potential is two-dimensional), $\theta_{\rm Ein}$ is scaled by the dynamical normalization factor $\lambda$, which is a function of the SIE axis ratio $q$.
To calculate this factor, the same approach as in OM10 \citep[and][equations 6,7, and 8]{Oguri2012} is followed.

The key properties of interest to microlensing are the values of the convergence, $\kappa$, and the shear, $\gamma$, at the locations of the multiple images.
These values are obtained from the following equations:
\begin{align}
\kappa(x,y) & = \frac{1}{2} \frac{\theta_{\rm Ein} \, \sqrt{q}}{\omega(x,y)}, \label{eq:k} \\
\gamma(x,y) & = | \vec{\gamma}^{\rm ext} + \vec{\gamma}^{\rm \scriptscriptstyle SIE}(x,y) |, \label{eq:g}
\end{align}
where $\omega(x,y) = \sqrt{q^2 x^2 + y^2}$ is the elliptical radius, and $x,y$ are the image coordinates in the frame rotated by the position angle of the ellipsoid, $\phi_{\rm L}$, so that its major axis aligns with the x-axis.
In this notation, $\theta_{\rm Ein}$ is measured on the intermediate (or equivalent) axis so that any integral of equation (\ref{eq:k}) (and equation \ref{eq:sersic}) within an ellipse would eventually be independent of the axis ratio $q$.

The shear vector is the sum of a component due to the lens potential (intrinsic):
\begin{equation}
\label{eq:gamma-sie-components}
\begin{split}
\gamma_1^{\rm \scriptscriptstyle SIE} & = (x^2-y^2) \, \frac{\kappa(x,y)}{x^2+y^2}, \\
\gamma_2^{\rm \scriptscriptstyle SIE} & = 2 \, x \, y \, \frac{\kappa(x,y)}{x^2+y^2}, \\
\end{split}
\end{equation}
with $x,y$ the same as before, and an external component due to environmental effects (e.g. nearby galaxies, local over/under-density, etc):
\begin{equation}
\label{eq:gamma-ext-components}
\begin{split}
\gamma_1^{\rm ext} & = \gamma^{\rm ext} \, \mathrm{cos}(2\phi_{\rm \gamma}), \\
\gamma_2^{\rm ext} & = \gamma^{\rm ext} \, \mathrm{sin}(2\phi_{\rm \gamma}). \\
\end{split}
\end{equation}
The magnitude of the external shear, $\gamma^{\rm ext}$, is assumed to follow a log-normal distribution with mean 0.05 and dispersion 0.2 dex (see OM10), and the orientation, $\phi_{\rm \gamma}$, is assumed to be random.
It can be shown that $\kappa = | \gamma^{\rm \scriptscriptstyle SIE} |$, as expected for an isothermal model.

\begin{figure}
	\includegraphics[width=0.5\textwidth]{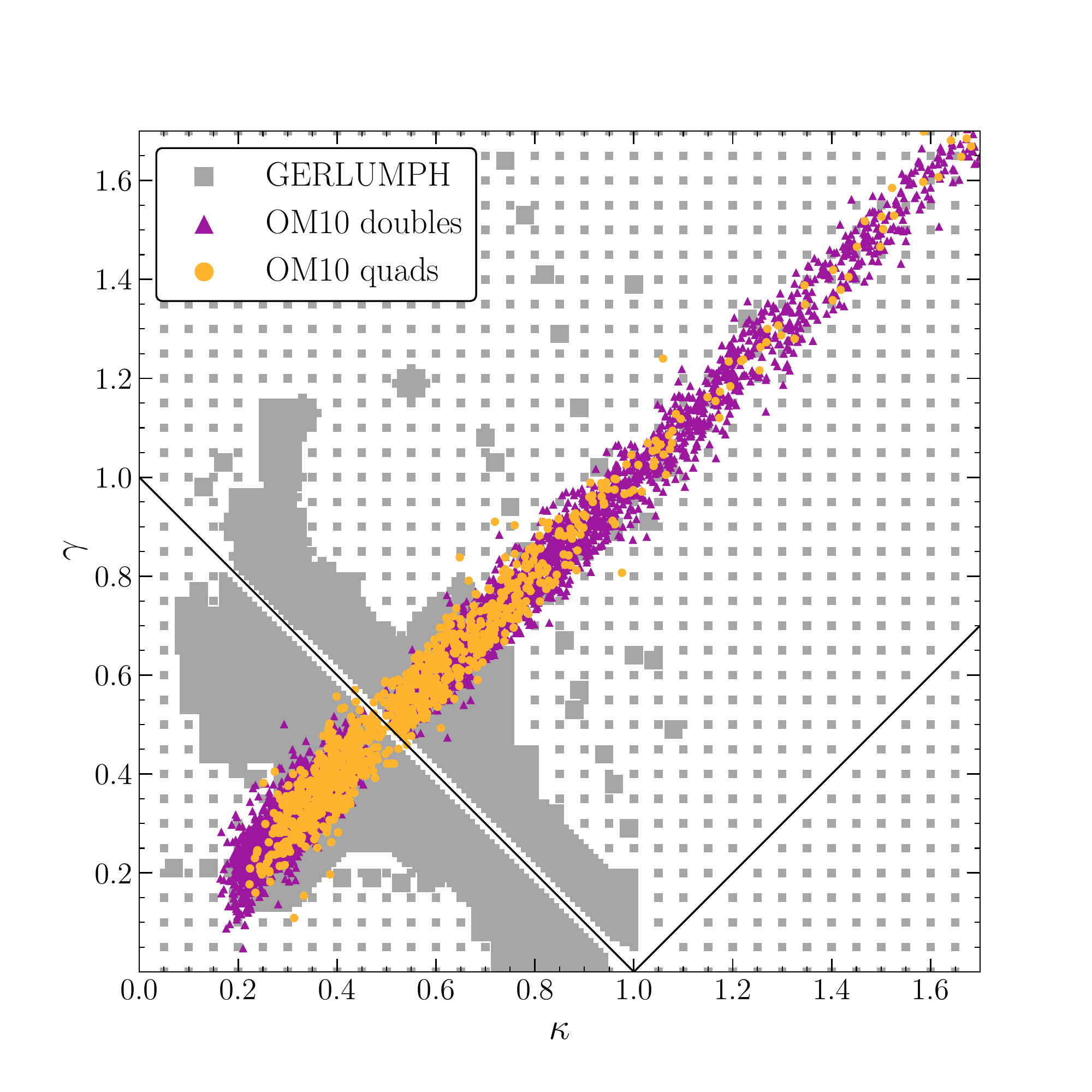}
	\caption{All the multiple images from the simulated 321 quadruply (circles, 1284 \kg locations) and 2025 doubly (triangles, 4050 \kg locations) imaged quasars plotted in the \kg parameter space. The critical line, i.e. the locus of points where $\mu \to\infty$ (equation \ref{eq:mag}), divides the parameter space in the minima (below the line, for $\kappa < 1$), saddle-point (above), and maxima (below, for $\kappa>1$) regions. The grey squares are the footprint of the GERLUMPH parameter survey \protect\citep[][]{Vernardos2014b,Vernardos2015}, and each one corresponds to maps with at least 11 different values of the smooth matter fraction $s$ (see text).}
	\label{fig:pspace}
\end{figure}

The image magnification without microlensing, or macro-magnification, can be obtained by:
\begin{equation}
\label{eq:mag}
 \mu = \frac{1}{(1-\kappa)^2-\gamma^2}.
\end{equation}
The image properties of the fiducial system shown in Fig. \ref{fig:example} are listed in Table \ref{tab:example_images}.
In Fig. \ref{fig:pspace}, the values of \kg are shown for the ensemble of lensed quasars selected from OM10 using the Euclid specifications.

\subsection{Smooth matter fraction}
\label{sec:smf}
An additional parameter required for the subsequent microlensing modelling is the smooth matter fraction:
\begin{equation}
\label{eq:smf}
s = 1 - \frac{\kappa_*}{\kappa},
\end{equation}
where $\kappa_*$ is the convergence due to matter in the form of compact (stellar) objects, and $\kappa$ is the total convergence given by equation (\ref{eq:k}).
The value of $s$, together with $\kappa,\gamma$, at the location of the multiple images are the main parameters for generating microlensing magnification maps (see below).

In order to obtain a value of $s$, a further assumption on the three-dimensional distribution of the stellar component of the lens needs to be made.
The Sersic model is adopted as descriptive of the stellar distribution of the elliptical lens galaxies.
This model is proven to successfully describe the main (bulge) component of virtually all early-type galaxies \citep[e.g.][]{Kormendy2016}.
Note that in this context the Sersic profile is adopted to define the radial mass distribution, rather than the projected surface brightness distribution.
Therefore, it is assumed that the mass radial profile is parametrized in a similar way to the light radial profile.
Deprojections of the Sersic profiles show that this is a valid assumption outside the innermost few parsec \citep[e.g. figure 1 of][]{Glass2011}.

An elliptical Sersic profile having the same orientation and ellipticity as the SIE is assumed:
\begin{equation}
\label{eq:sersic}
\kappa_{*,n} (x,y) = A_n \, \mathrm{exp}\{ \, \, k_n \left[ \frac{\omega(x,y)}{\sqrt{q} \, \theta_{\rm eff}} \right]^{1/n}\},
\end{equation}
where $n$ is the Sersic index, $k_n$ is a constant \citep[see][]{Capaccioli1989}, and $\omega$ and $q$ are the same as in equation (\ref{eq:k}), meaning that $\theta_{\rm eff}$ is the effective radius on the intermediate axis.
The normalization constant $A_n$ can be computed by matching the dark matter fraction, $f_{\rm DM}$, of the central region of a lensing galaxy ($< \theta_{\rm eff}/2$) with a total mass $M_{\rm tot}$ (both dark and baryonic) to observations \citep[see also][]{FoxleyMarrable2018}:
\begin{align}
\label{eq:norm}
A_n 	 	& =  (1-f_{\rm DM}) \frac{M_{\rm tot}}{Q_n},        \\
M_{\rm tot} & =  \int_0^{\theta_{\rm eff}/2} \kappa(x,y) \, \mathrm{d}x \, \mathrm{d}y, \label{eq:Mtot} \\
Q_n         & =  \int_0^{\theta_{\rm eff}/2} \mathrm{exp}\{ \, \, k_n \left[ \frac{\omega(x,y)}{\sqrt{q} \, \theta_{\rm eff}} \right]^{1/n}\} \, \mathrm{d}x \, \mathrm{d}y \label{eq:Q}.
\end{align}
The solutions to integrals (\ref{eq:Mtot}) and (\ref{eq:Q}) are $\pi \, \theta_{\rm Ein} \, \theta_{\rm eff}/2$ and $2 \pi \, n \, \theta_{\rm eff}^2 \, k_n^{-2n} \, \gamma(2n,\xi_0)$, respectively, where $\gamma(2n,\xi_0)$ is the incomplete gamma function, and $\xi_0 = k_n/2^{1/n}$.
\citet{Auger2010} have provided fits of the central dark matter fraction for the Sloan Lens ACS (SLACS) sample of strong elliptical lenses of the form:
\begin{equation}
\label{eq:imf}
f_{\rm DM} = \alpha \, \mathrm{log}_{10} \left( \frac{\upsilon'}{100 \, \mathrm{km \, s^{-1}}} \right) + \beta,
\end{equation}
where $\upsilon'$ is the velocity dispersion within $\theta_{\rm eff}/2$, which is effectively equal to $\upsilon_{\rm disp}$ measured at the center of the lensing galaxy \citep[see equation 2 of][]{Jorgensen1995}.
They find $\alpha=0.46\pm0.22,\beta=0.40\pm0.09$ for a Chabrier, and $\alpha=0.8\pm0.44,\beta=-0.05\pm0.18$ for a Salpeter Initial Mass Function (IMF).

\begin{figure}
	\includegraphics[width=0.47\textwidth]{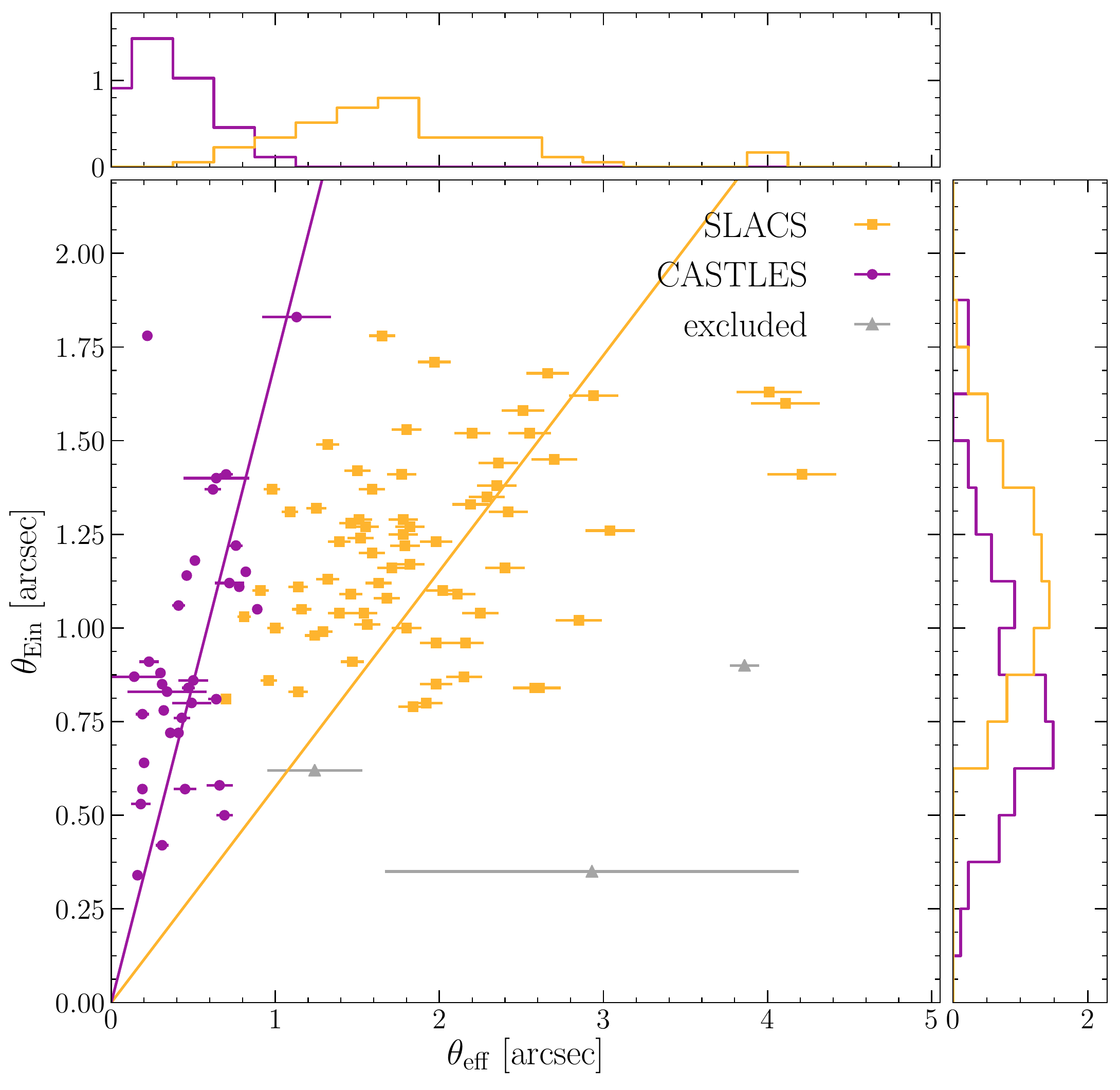}
	\caption{Measured effective and Einstein radii for the SLACS \protect\citep[squares,][]{Auger2009} and the CASTLES \protect\citep[circles,][]{Oguri2014} lenses (main panel). The lines represent linear fits of the form $\theta_{\rm Ein} = a \, \theta_{\rm eff}$, with $a=0.576$ and $a=1.71$ for each sample respectively. The triangles indicate outliers that were excluded from the fit for the CASTLES sample: Q2237$+$030 (with the smallest errorbars), the usual non-typical quadruple lens (the lens lies at the exceptionally low redshift of 0.04), and two double lenses, Q1355-2257 and FBQ1633$+$3134, whose lens galaxy light is very faint to provide a well measured effective radius. The histograms indicate the probability densities for $\theta_{\rm eff}$ (top panel) and $\theta_{\rm Ein}$ (right panel).}
	\label{fig:reff_rein}
\end{figure}

To assign a value to $\theta_{\rm eff}$, measured values from two different samples of observed gravitational lenses are used.
\citet{Auger2009} estimate $\langle \theta_{\rm Ein}/\theta_{\rm eff} \rangle = 0.576$ for the SLACS lenses.
However, SLACS lenses are strong galaxy-galaxy lenses, while the CASTLES sample of lensed quasars can be considered more suitable for the mock systems studied in this work.
\citet{Oguri2014} have modelled the light and mass profiles for the CASTLES systems deriving $\langle \theta_{\rm Ein}/\theta_{\rm eff} \rangle = 1.71$.
The data and linear fits for the SLACS and CASTLES samples are shown in Fig. \ref{fig:reff_rein}, and the source and lens redshift distributions in Fig. \ref{fig:z_hist}.

\begin{figure}
	\includegraphics[width=0.47\textwidth]{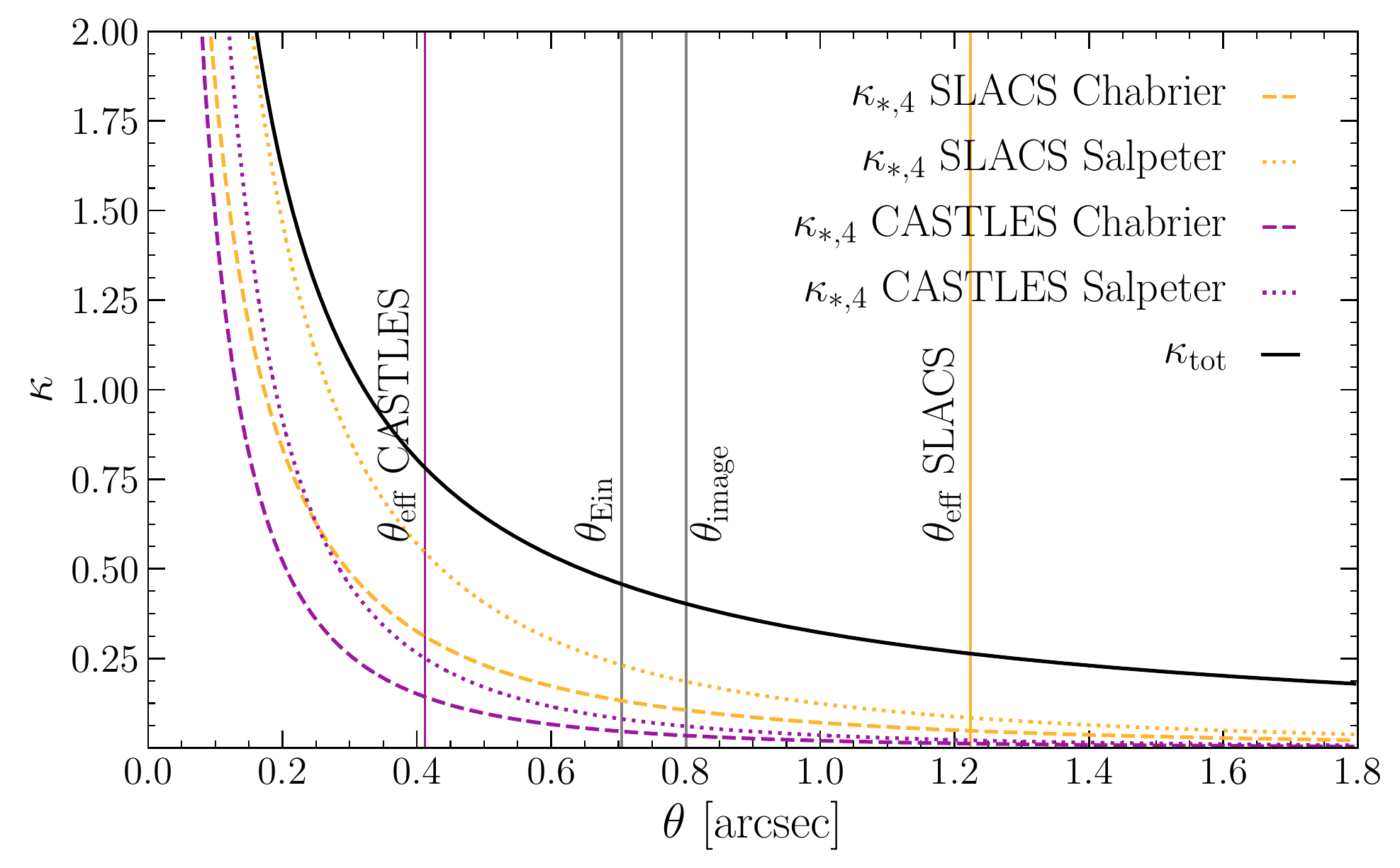}
	\caption{Convergence profiles for image C of the fiducial system of Fig. \ref{fig:example}. These are one-dimensional de Vaucouleurs ($n=4$) profiles extracted along the direction from the center of the lens to the location of the image. The total convergence, $\kappa_{\rm tot}$, of the corresponding SIE mass profile is also shown. The different compact matter profiles, $\kappa_{*,4}$, have been calibrated using a Chabrier or Salpeter IMF and the SLACS or CASTLES $\theta_{\rm eff}/\theta_{\rm Ein}$ relation (see Fig. \ref{fig:reff_rein}). The locations of the image, Einstein radius (tangential critical line), and the different effective radii, are indicated by the labeled vertical lines. The resulting smooth matter fractions at the image location are shown in Table \ref{tab:example_smf}.}
	\label{fig:example_profiles}
\end{figure}

Assigning the value of $\theta_{\rm eff}$ from these two different lens samples, and assuming different IMFs to calculate the normalization of the stellar matter component of equation (\ref{eq:sersic}), leads to different values of $s$ at the locations of the multiple images.
An example is shown in Fig. \ref{fig:example_profiles} for image C of the fiducial system of Fig. \ref{fig:example}, adopting de Vaucouleurs profiles for $\kappa_*$ ($n=4$).
The $s$ values for all images are shown in Table \ref{tab:example_smf}.

\begin{figure}
	\includegraphics[width=0.47\textwidth]{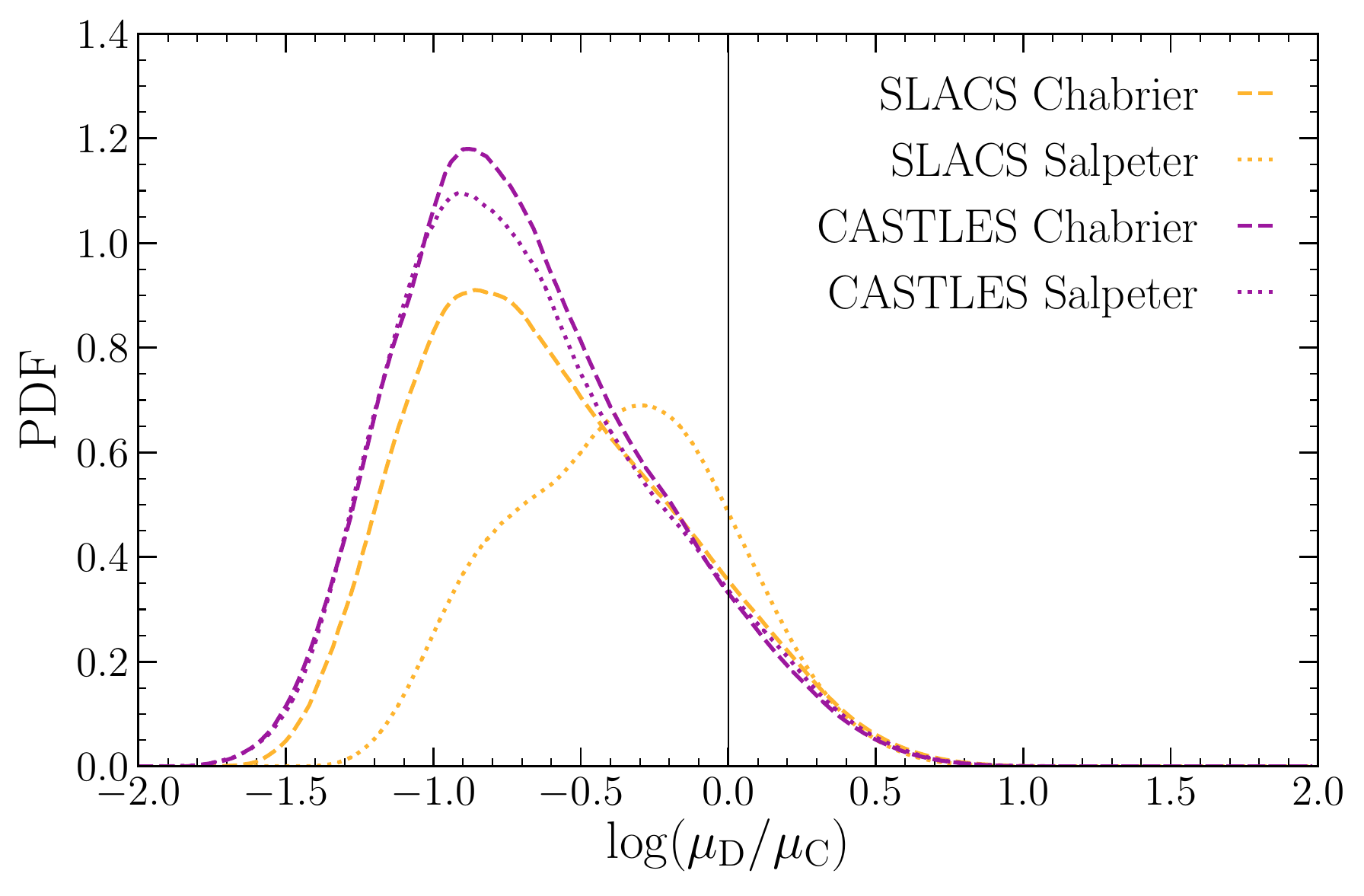}
	\caption{Simulated flux ratio probability density between images D (saddle-point) and C (minimum) of the fiducial system of Fig. \ref{fig:example}, for a SLACS- or CASTLES-like sample of lenses, having a Salpeter or Chabrier IMF. An accretion disc with $\nu = 1$ and $r_0=2.59\times10^{15}$ cm was assumed, observed at $\lambda_{\rm obs} = 550$. The vertical line indicates the value of the ratio in the absence of microlensing (in this and all the subsequent similar plots, the microlensing magnifications have been divided by the macro-magnification from equation \ref{eq:mag}).}
	\label{fig:example_rpds}
\end{figure}

\begin{table}
	\centering
	\caption{Smooth matter fraction values for the multiple images of the fiducial system of Fig. \ref{fig:example}. Values are shown for a $\theta_{\rm eff}/\theta_{\rm Ein}$ relation extracted from the SLACS and CASTLES samples (see Fig. \ref{fig:reff_rein}), using either a Chabrier or a Salpeter IMF.}
	\label{tab:example_smf}
	\begin{tabular}{rcccc}
							& A		& B		& C		& D 	\\
		\hline
		SLACS, Chabrier		& 0.750	& 0.625	& 0.719	& 0.666 \\
		SLACS, Salpeter		& 0.562	& 0.343	& 0.508	& 0.415 \\		
		CASTLES, Chabrier	& 0.921	& 0.837	& 0.904	& 0.869 \\		
		CASTLES, Salpeter	& 0.862	& 0.715	& 0.831	& 0.770 \\		
		\hline
	\end{tabular}
\end{table}

\subsection{Source profile}
\label{sec:source}
The relevant source component for this microlensing study is the quasar accretion disc; microlensing of the broad emission line region or of any other structure in the source is disregarded.
Microlensing induces brightness variations depending on the size of the source relative to the size of the microlensing caustics and their clustering.
In general, the smaller the source with respect to the Einstein radius of the microlenses on the source plane (see equation \ref{eq:rein}) the larger the (de)magnification due to microlensing can be.

The size of the accretion disc as a function of wavelength can be described by a parametric model, e.g. the following power law:
\begin{equation}
\label{eq:parametrized}
r = r_0 \left(\frac{\lambda}{\lambda_0}\right)^{\nu},
\end{equation}
where $\lambda = \lambda_{\rm obs}/(1+z_S)$, with $\lambda_{\rm obs}$ being the observing wavelength, $z_S$ the source redshift, $\nu$ the power law index, and $r_0$ the size of the disc observed at the rest wavelength $\lambda_0 = 102.68$ nm \citep[see e.g.][]{Jimenez2014,Bate2018}.
Formally, a specific accretion disc model \citep[e.g. the thin disc model, see equation 9 of][]{Vernardos2014c} may be used.
However, due to the degeneracies explained below, the more general parametric form of equation (\ref{eq:parametrized}) is deemed sufficient for the purposes of this work.

The ratio of the source size at a given observed wavelength (equation \ref{eq:parametrized}) over the Einstein radius (equation \ref{eq:rein}), $r/R_{\rm Ein}$, is an indicator of the strength of microlensing effects.
This ratio depends on the accretion disc profile parameters $r_0$ and $\nu$, and on the redshifts of the lens and source in a non-linear way.
However, it is pointed out that these parameters are degenerate.
In practice, assuming a population of lenses with a given distribution in $z_S$ and the same accretion disc (size and shape, i.e. keeping $r_0$ and $\nu$ fixed) has an equivalent effect on $r/R_{\rm Ein}$ as assuming a population of lenses having the same source and lens redshifts but varying $r_0$ and/or $\nu$ for the accretion disc.
For example, assuming the same accretion disc with $\nu=1$ and $r_0 = 2.59 \times 10^{15}$cm (or 1 light day) for all the lenses in the OM10 sample gives $0.058 < r/R_{\rm Ein} < 0.42$ at $\lambda_{\rm obs} = 900$ nm.
The same variation of $r/R_{\rm Ein}$ - and hence the same microlensing effect - can be produced by assuming a fixed lens and source redshift of $z_L=0.5$ and $z_S=2.0$ ($R_{\rm Ein} = 5.35 \times 10^{16}$cm for 1 M$_{\odot}$ microlenses), and a population of accretion discs with $10^{15}$cm $< r_0 < 7.7 \times 10^{15}$cm, or $0.06 < \nu < 0.74$.

The focus of this paper is to examine the microlensing observables as a function of the mock lens properties, and \emph{not} to derive the properties of the source.
A fiducial accretion disc profile with $\nu=1$ and $r_0 = 2.59 \times 10^{15}$cm (1 light-day) is used throughout the simulations presented here\footnote{The exponent $\nu$ is mostly of interest when flux ratios in different wavelengths are examined, e.g. to determine the accretion disc temperature profile, which is not the case here (see Section \ref{sec:microlensing}). However, its value, together with $r_0$, sets the size of the accretion disc in any given wavelength via equation (\ref{eq:parametrized}).}.
This value of $r_0$ is roughly equal to ten times the gravitational radius of a typical $10^9$ M$_{\odot}$ black hole and five times the size of a corresponding thin disc model\footnote{The thin disc model \citep{Shakura1973} predicts an accretion disc size of:
\begin{equation*}
r_{0,thin} = 3.75 \, \left( \frac{M_{\rm BH}}{10^9 \mathrm{M}_{\odot}} \right)^{2/3} \left( \frac{f_{\rm Edd}}{\eta} \right)^{1/3} \left( \frac{\lambda_{\rm rest}}{\mu m} \right)^{4/3},
\end{equation*}
measured in light-days.
Using $\eta=0.15$ and $f_{\rm Edd} = 0.25$ as typical values for the accretion efficiency and the Eddington ratio \citep[e.g.][]{Vernardos2014c,Blackburne2011}, a black hole mass of $10^9$ M$_{\odot}$, and $\lambda_{\rm rest} = \lambda_0 = 102.68$ nm (as in equation \ref{eq:parametrized}) results to a size of $r_{0,thin} \approx 0.21$ light-days for the accretion disc.}.
However, microlensing studies seem to find a somewhat higher value of $r_0 = 4.5$ light-days \citep[e.g. see][who study a collection of 12, 10, and 4 systems respectively]{Blackburne2011,Jimenez2014,Bate2018}, which is larger than what is expected from standard accretion disc theory.
It is still unclear whether this is due to the small number of objects considered in each case, inappropriate theoretical treatment of quasar accretion discs, or caveats in the microlensing techniques used \citep[as suggested recently by][for the power law dependence of the size on the wavelength]{Bate2018}.
Therefore, both values of $r_0$, 1 and 4.5 light-days respectively, are used in the case of doubly imaged quasars (see Section \ref{sec:doubles}), where the former value is the fiducial value used in the rest of the simulations without any prior information on the source size.

The lensed quasars selected from OM10 (Section \ref{sec:mocks}) are already constraining the possible accretion disc profiles in that they are biased towards luminous objects.
Additionally, the accretion disc profiles can be assumed to evolve with redshift.
Nevertheless, neither constraints on the luminosity nor evolution with redshift of the properties of the source are assumed in the following and the problem is tackled more broadly at the cost of some additional computations due to the larger parameter space explored.
The focus remains on the lensing galaxies and recovering properties of the source is left for future work.

\subsection{Microlensing}
\label{sec:microlensing}
An ensemble of stellar-mass microlenses creates a network of caustics in the source plane that is best described by a pixellated magnification map \citep[see][]{Kayser1986}.
The main parameters of such a map are $\kappa,\gamma$, and $s$, while its width is measured in units of the projected Einstein radius of a point-mass microlens on the source plane:
\begin{equation}
\label{eq:rein}
R_{\rm Ein} = \sqrt{\frac{4G\langle M \rangle}{c^2} \frac{D_{\mathrm{S}} \, D_{\mathrm{LS}}}{\mathrm{D_L}}},
\end{equation}
with $D_{\mathrm{S}},D_{\mathrm{L}},$ and $D_{\mathrm{LS}}$ being the angular diameter distances to the source, the lens, and between the lens and the source respectively, $M$ the microlens mass, $G$ the gravitational constant, and $c$ the speed of light.

For general populations of simulated lenses [i.e. not restricted to the \kg values of the sample used here (see Fig. \ref{fig:pspace}) nor to the approach followed in assigning a value for $s$] the precomputed available GERLUMPH magnification maps can be used.
These square maps have a width of 25 $R_{\mathrm{Ein}}$ and a resolution of 10,000 on each side.
For each \kg combination, the GERLUMPH maps cover $0 \leq s \leq 0.9$ in steps of 0.1, plus $s=0.95, 0.99$.
To adopt the proper magnification map (specified by the given $\kappa,\gamma$, and $s$), the set of maps generated by GERLUMPH is traversed and the map with $\kappa,\gamma,$ and $s$ values closest to the ones searched for is selected.
In all cases, the difference between the actual and the matched GERLUMPH values is $\Delta \kappa,\Delta \gamma < 0.05$ and $s<0.1$.

\begin{figure*}
	\includegraphics[width=\textwidth]{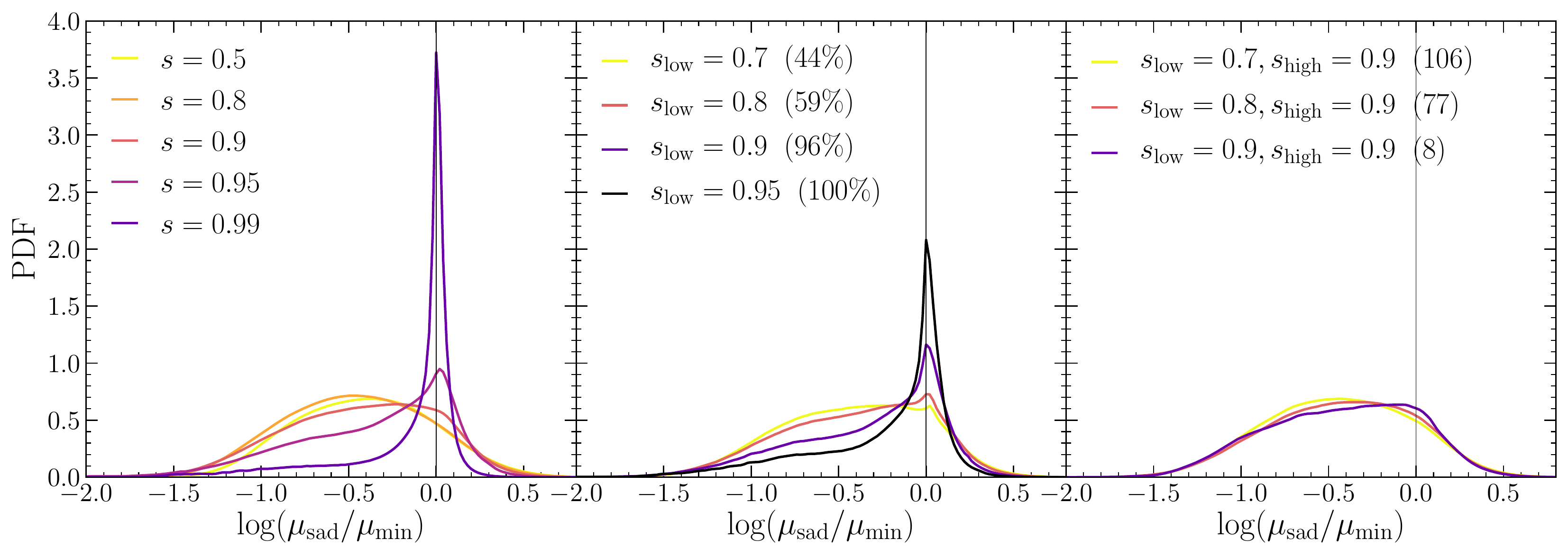}
	\caption{Flux ratio probability density per system for collections of close image pairs under different assumptions on setting the value of $s$ for each image. The microlensing magnifications have been normalized by the corresponding macro-magnifications. The vertical line indicates the value of the ratio in the absence of microlensing. Left: A sample of 188 close image pairs, with both images having the same value of $s$. Middle: The same sample of 188 close image pairs, with $s$ assigned randomly in the range $[s_{\rm low},0.99]$ for each image. Numbers in parenthesis indicate the percentage of pairs in the sample with at least one image having $s>0.9$. Right: Same as the middle panel, excluding all the pairs having at least one image with $s>0.9$. Numbers in parenthesis indicate how many image pairs remain in the sample in each case.}
	\label{fig:result_constants_semirandom}
\end{figure*}

\begin{figure*}
	\includegraphics[width=\textwidth]{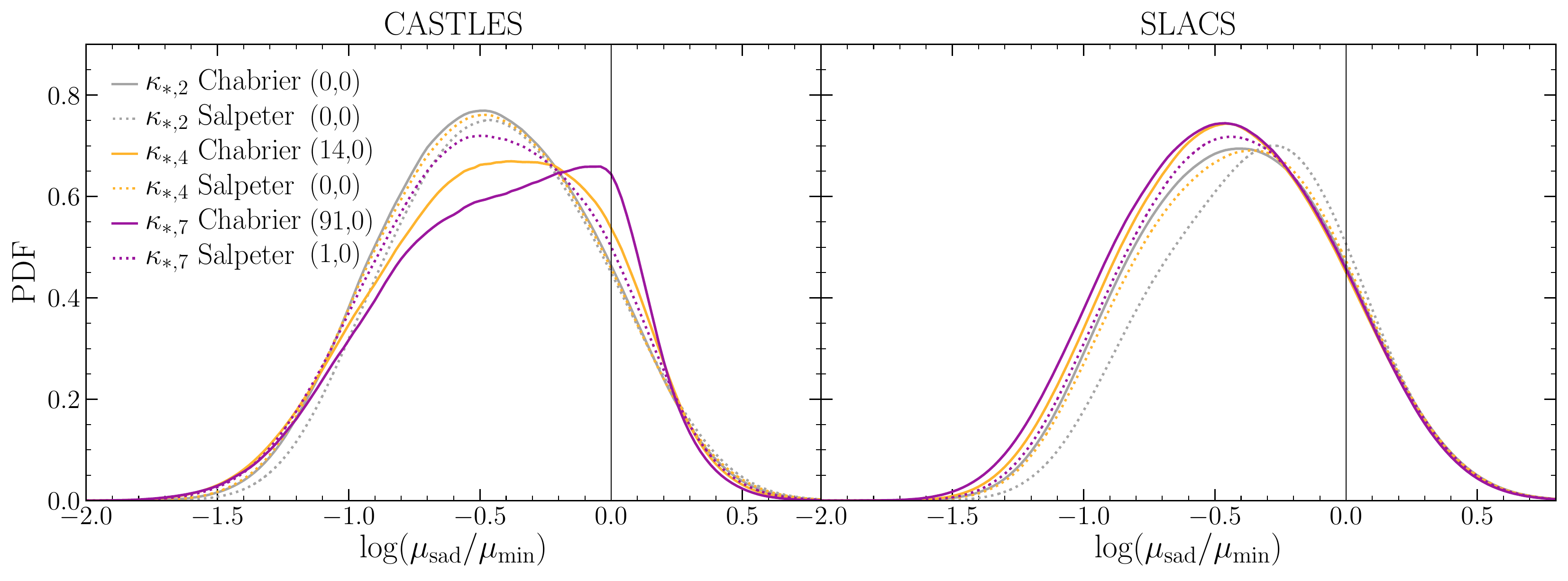}
	\caption{Flux ratio probability density per system for mock lensed populations having the CASTLES (left) or SLACS (right) $\theta_{\rm Ein}/\theta_{\rm eff}$ relation. Each collection of systems has the same IMF and Sersic index $n$ (equation \ref{eq:sersic}, indicated as a subscript for $\kappa_*$) for the compact matter distribution. Solid (dotted) lines correspond to a Chabrier (Salpeter) IMF and different colors (grey, yellow, purple) to different Sersic indices $n$ (2, 4, 7). The microlensing magnifications have been divided by the macro-magnification for each image. The vertical line indicates the value of the ratio in the absence of microlensing. The numbers in the parenthesis indicate the percentage of image pairs with at least one image having $s=0.95$ for the CASTLES and SLACS cases respectively.}
	\label{fig:result_realistic_closest}
\end{figure*}

For a finite source having a size greater than the pixel size of the magnification map (this is $0.0025 \, R_{\rm Ein}$ for the GERLUMPH maps), a convolution needs to be performed between the map and the source profile \citep[e.g.][]{Bate2011,Jimenez2014,Bate2018,Vernardos2018a}.
The shape\footnote{As \citet{Mortonson2005} have shown, the microlensing effects are insensitive to the actual shape of the source profile, while crucially depending on its size.} of the source profile is assumed to be a symmetric (face-on) two-dimensional Gaussian, truncated at twice the half-light radius, $r_{1/2}$, where $r_{1/2} = 1.18 \sigma$, and the value of $\sigma$ (the standard deviation of the Gaussian) is set to be equal to $r_0$ from equation (\ref{eq:parametrized}).
Each convolved map is then reduced to an effective convolved map, i.e. a downsized version of the convolved map excluding the edge effects produced by the convolution with the source profile.

Mock flux ratios can be obtained after getting the ratio of magnifications of pixels sampled from a pair of effective convolved maps.
While sampling from an effective map, one has to take into account two effects: a large number of pixels is required to correctly reproduce the underlying magnification probability distribution, and these pixels should be away from each other in order to be statistically independent (the convolution introduces correlations to the pixels at the scale of the source size).
To compensate for these two effects, a total of $10^4$ pixels are selected from a 100x100 regular grid of locations covering the whole effective map.
The resulting sets of pixels between two maps are subsequently divided to produce a set of $10^8$ mock flux ratios.
The probability distribution of such mock flux ratios is shown in Fig. \ref{fig:example_rpds} for the close pair of images C and D in the example system of Fig. \ref{fig:example}.
Such probability distributions are interesting in their own right, e.g. they can be useful in estimating the most probable flux ratio due to microlensing in different wavelengths.
In Section \ref{sec:results} the combined PDFs of different populations of mock quasars are examined as a whole.

\section{Results}
\label{sec:results}
In this section, the aggregate theoretical flux ratio probability distribution of a collection of mock lensed quasars is computed under different assumptions for the lens mass components.
The microlensing magnifications are always normalized to the macro-magnification.
In this way, a ratio of unity between the magnifications observed in some fiducial images A and B corresponds to $\mu_{\rm A}/\mu_{\rm B}$, where $\mu$ is given by equation \ref{eq:mag}.
In the following, several scenarios are examined starting with simple `toy' models for $s$, and then using more realistic ones based on the IMF (Chabrier or Salpeter) and the $\theta_{\rm Ein}/\theta_{\rm eff}$ relation (see Fig. \ref{fig:reff_rein} and related text).
The reader is warned about two caveats to be considered in evaluating the results presented here.

One caveat regards the fact that the Euclid VIS instrument covers a wide range of wavelengths: 550nm $< \lambda_{\rm obs} <$ 900nm.
In order to avoid accounting for the structure of the source over such a broad range of wavelengths - which implies considering additional contamination by other non-microlensed features - the mock observations are assumed to be at $\lambda_{\rm obs} = 900$nm.
This sets a lower limit to the microlensing effects.
Although assuming observations at shorter wavelengths (e.g. $\lambda_{\rm obs} = 550$nm, at which the source should appear more compact) is expected to enhance the microlensing effect, this is not observed in the results presented below. 
However, wavelengths longer than 900nm would lead to larger apparent accretion discs and gradually wash out any microlensing induced deviation from the macro-magnification.
For example, the PDFs in Fig. \ref{fig:example_rpds}, would peak around $\mu_{\rm sad}/\mu_{\rm min} = 1.0$ and become narrower, regardless of the IMF, the $\theta_{\rm Ein}/\theta_{\rm eff}$ relation, or the Sersic index of the compact matter profile used.
Hence, all the mock observations presented below are valid for the long wavelength end of the Euclid VIS instrument, and no other wavelengths are considered.

The second caveat regards image pairs.
Close image pairs in quadruple image configurations are produced by the quasar lying near a fold caustic in the source plane.
The magnification in such observed image pairs is less dependent on the macro-model and therefore deviations from it (anomalies) are more easily detected \citep[e.g. see][]{Gaudi2002a}.
Nevertheless, such anomalies manifest themselves independently of the macro-image configuration due to their origin, e.g. existence of higher-order moments in the lens mass distribution, or presence of massive substructures.
These phenomena are therefore of great interest and were carefully simulated in the following analysis.
In order to construct a sample consisting of close image pairs (fold geometries), images separated by a factor of $<0.8$ than that of the next closest pair are selected.
In this way, symmetric cross-like (distances between all images are almost the same) and cusp-like (three images lying close to each other in roughly the same distance along an arc) image configurations are excluded.
Selecting tighter pairs (by reducing the separation factor) leads to identical results as below, but also to smaller samples of objects.
The final sample consists of $188$ close image pairs.

\subsection{Simple models for compact matter}
As a first test, $s$ is set to be the same for both images in the pair.
Flux ratio probability distributions are produced for each lens as described in Section \ref{sec:microlensing} and shown in Fig. \ref{fig:example_rpds}, and then added to create the aggregate flux ratio probability distribution for the entire sample.
In this context, the lenses are treated as an unlabelled set.
Factoring out the actual number of objects one retrieves the probability density function per system, PDF, shown in the panels of Fig. \ref{fig:result_constants_semirandom}.
The left panel of Fig. \ref{fig:result_constants_semirandom} shows the PDFs for different values of constant $s$.
For low values ($s \leq 0.9$) the sample probability density is dominated by a broad peak at $\mu_{\rm sad}/\mu_{\rm min} \approx 0.3$ [$\mathrm{log}(\mu_{\rm sad}/\mu_{\rm min}) \approx -0.5$].
As $s$ increases there is a fast transition in the shape of the distribution in the range $0.9<s<0.95$, which is now dominated by an emergent sharp peak at $\mu_{\rm sad}/\mu_{\rm min} = 1.0$, consistent with no microlensing taking place.
This is to be expected since high values of $s$ mean less matter in the form of compact microlenses and the individual magnification probability distributions for each image are concentrated tightly around the macro-magnification \citep[see e.g. fig. 3 in][]{Vernardos2014a}.

Another test was performed this time assigning a random value to $s$ in the range $[s_{\rm low},0.99]$ for the images.
The resulting PDFs are shown in the middle panel of Fig. \ref{fig:result_constants_semirandom}, and demonstrate that both of the PDF peaks presented above, at $\mu_{\rm sad}/\mu_{\rm min} \approx 0.3$ and $\mu_{\rm sad}/\mu_{\rm min} = 1.0$ are in fact present albeit having varying amplitudes.
The amplitude of the peak around $\mu_{\rm sad}/\mu_{\rm min} = 1.0$ is due to the percentage of pairs within the sample with at least one of the images having $s>0.9$ (either $s=0.95$ or $s=0.99$).
This peak will disappear if such pairs are excluded, as shown in the right panel of Fig. \ref{fig:result_constants_semirandom}.

\subsection{Realistic models}
How will the PDFs look if realistic assumptions are made for the compact matter distribution within the lensing galaxy?
Assuming a specific IMF, a Sersic index $n$ (equation \ref{eq:sersic}) for the compact matter profile, and a $\theta_{\rm Ein}/\theta_{\rm eff}$ relation for the mock lenses results in different values of $s$ for the images.
The same procedure described above is followed to create the sample PDFs.

From the results presented in Fig. \ref{fig:result_constants_semirandom}, it is generally expected that the less the amount of compact matter (higher $s$) at the locations of the close image pairs, the more prominent the peak of the PDFs around $\mu_{\rm sad}/\mu_{\rm min} = 1.0$ will be.
On top of this, the PDFs will have a broad peak around $\mu_{\rm sad}/\mu_{\rm min} = 0.3$.
For configurations with a dominant compact matter component, this will be the only visible peak.
For example, a higher Sersic index is expected to lead to a steeper compact matter profile, a higher value of $s$ at the location of the images, and a PDF with a dominant peak at $\mu_{\rm sad}/\mu_{\rm min} = 1.0$, and vice versa.
In Fig. \ref{fig:result_realistic_closest}, the PDFs of a sample of 188 close image pairs are shown under different assumptions for the IMF, the $\theta_{\rm Ein}/\theta_{\rm eff}$ relation, and the Sersic index $n$.

Focusing on the left panel of Fig. \ref{fig:result_realistic_closest}, which shows lens populations following the $\theta_{\rm Ein}/\theta_{\rm eff}$ relation from CASTLES, the prominence of the peak at $\mu_{\rm sad}/\mu_{\rm min} = 1.0$ is again dependent on the number of images with $s>0.9$.
It turns out that only the combination of a Sersic index of $n=7$ and a Chabrier IMF results in a high number of pairs with at least one such image: $91\%$ of the 188 pairs (see numbers in parenthesis in left panel of Fig \ref{fig:result_realistic_closest}).
Although this percentage is high, the prominence of the peak is still inferior to those of the cases shown in the middle panel of Fig. \ref{fig:result_constants_semirandom} due to the fact that all the images with $s>0.9$ actually have $s=0.95$, none of them having $s=0.99$.

In the right panel of Fig. \ref{fig:result_realistic_closest}, the $\theta_{\rm Ein}/\theta_{\rm eff}$ relation from SLACS is used for the mock lenses.
For such systems, $\theta_{\rm eff}$ is in general larger, meaning that there is more light from the lens galaxy present at the location of the multiple quasar images.
In turn, a higher number of stars is required to generate this light, leading to higher $\kappa_*$ and lower $s$, and there is no combination of IMF and Sersic index from the ones used here that can produce any image with $s>0.9$.
Consequently, the PDFs in the right panel of Fig. \ref{fig:result_realistic_closest} have only one broad peak around $\mu_{\rm sad}/\mu_{\rm min} = 0.3$.

\begin{figure}
	\includegraphics[width=0.47\textwidth]{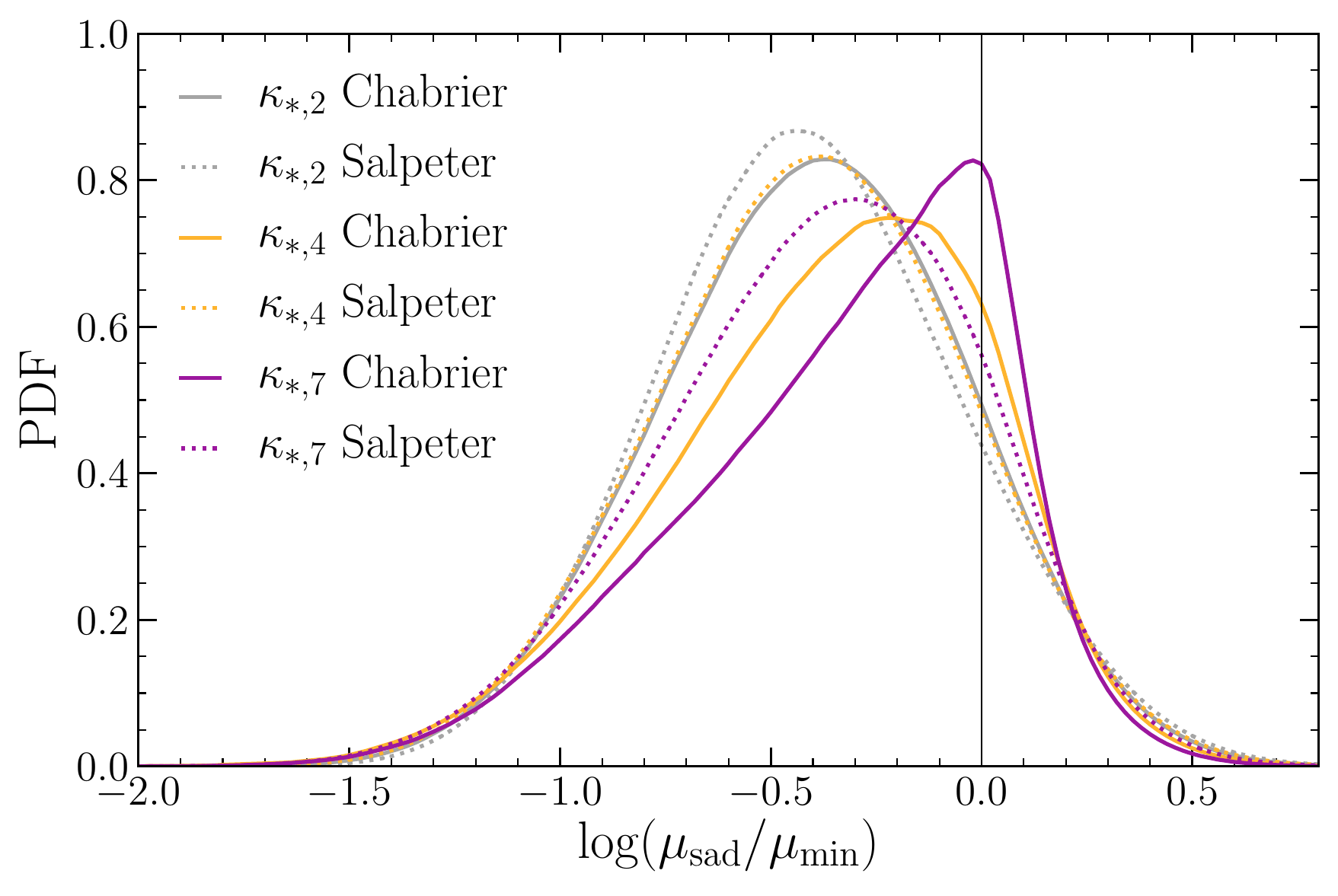}
	\caption{Same as left panel of Fig. \ref{fig:result_realistic_closest}, using the flux ratios from all the possible image pairs consisting of a saddle-point and a minimum. The mock lenses are assumed to follow the $\theta_{\rm Ein}/\theta_{\rm eff}$ relation from CASTLES.}
	\label{fig:result_realistic_all}
\end{figure}

\subsection{Using more images}
So far, only the closest image pairs from quadruple lenses in fold image configurations were used.
These pairs are expected to be generally insensitive to the actual lens model, making them the easiest targets to apply the current analysis to.
However, other image pairs will in general have different properties and expected PDFs; e.g. images A and B of the fiducial system shown in Fig. \ref{fig:example} are lying at different locations within the elliptical lens' total and compact mass profiles (closer and further, respectively, than the merging image pair) and will therefore have different $\kappa,\gamma,$ and $s$, properties than images C and D.
Including such other pairs in the analysis leads to different PDFs.

Dropping the requirement for a fold geometry, the number of retrieved lenses increases to 261\footnote{Due to some lens geometries from the OM10 catalog being `problematic', e.g. not producing exactly 2 minima and 2 saddle-points, and because the (\kg) values for some images lie outside the GERLUMPH range, this number is not equal to 321, i.e. the number of lenses predicted for the Euclid wide survey. The same holds for the case of doubly imaged quasars examined in Section \ref{sec:doubles}.}.
For each pair of images examined, the flux ratio PDF is calculated as explained in Section \ref{sec:microlensing} (see also Fig. \ref{fig:example_rpds}).
The aggregate PDF per image pair of a population of mock lenses is then obtained as previously by summing the individual PDFs.

Fig. \ref{fig:result_realistic_all} is the same as the left panel of Fig. \ref{fig:result_realistic_closest}, except that multiple image pairs have been used to create the PDFs and not just the closest one.
In this case, all the possible pairs between a saddle-point and a minimum image have been used (4 pairs in total for each lens).
From the several other pair combinations that were tested, it turned out that this one produces the most easily distinguishable differences to the PDFs.

As pointed out before, the larger amount of smooth matter a lens contains at the locations of its multiple images the stronger the peak at $\mu_{\rm sad}/\mu_{\rm min} = 1.0$.
This is also observed Fig. \ref{fig:result_realistic_all}, where a Chabrier IMF and a Sersic index $n=7$ lead to higher $s$ with a more peaked distribution with respect to Fig. \ref{fig:result_realistic_closest}.
Using the $\theta_{\rm Ein}/\theta_{\rm eff}$ relation from SLACS instead of CASTLES leads to PDFs that have very similar shapes to each other and cannot be easily distinguished.

\subsection{Doubly imaged quasars}
\label{sec:doubles}
In general, doubly imaged quasars provide less constraints to the lens and the source, e.g. there is only a single flux ratio between the images that can be measured.
This single flux ratio is sensitive to the exact macro-magnification and macromodel; uncertainties in the lens or line-of-sight substructure and deviations of the macromodel from the SIE could mask themselves as microlensing.
Nevertheless, doubly lensed quasars are more abundant by almost an order of magnitude with respect to the quadruple ones.
Therefore, even though such systems may be less constrained, their large numbers can still be used to study the lens mass configuration.

\begin{figure*}
	\includegraphics[width=\textwidth]{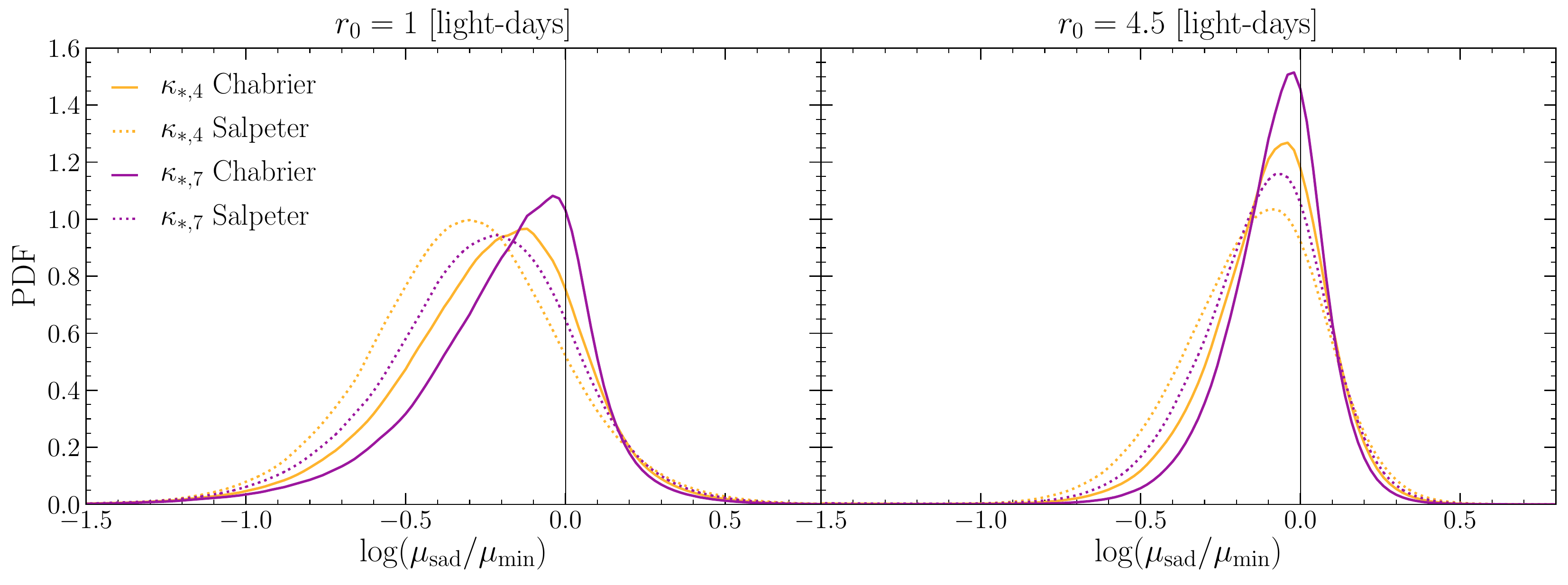}
	\caption{Same as Fig. \ref{fig:result_realistic_all} and left panel of Fig. \ref{fig:result_realistic_closest}, using the images from doubly lensed quasars. The mock lenses are assumed to follow the $\theta_{\rm Ein}/\theta_{\rm eff}$ relation from CASTLES. Two populations of accretion discs are examined using equation (\ref{eq:parametrized}) with $r_0$ equal to 1 (left panel) and 4.5 (right panel) light-days.}
	\label{fig:result_doubles_pdf}
\end{figure*}

The left panel of Fig. \ref{fig:result_doubles_pdf} shows the PDF for 1774 doubly imaged quasars under different assumptions for the IMF and the Sersic index of the compact matter distribution (the mock quadruple lens populations with $n=2$ and $n=4$ have quite similar properties, e.g. see Figs. \ref{fig:result_realistic_closest} and \ref{fig:result_realistic_all}, therefore the $n=2$ case is omitted in the simulated populations of doubles).
For this set of mocks, the $\theta_{\rm Ein}/\theta_{\rm eff}$ relation from CASTLES was used, which produces obvious differences in the PDFs as opposed to the one from SLACS.
Once again, collections of image pairs with higher smooth matter fractions are expected to have PDFs that peak closer to $\mu_{\rm sad}/\mu_{\rm min} = 1.0$.
This is the case for a Chabrier IMF with a Sersic index $n=7$.

The same mock population of doubles was produced, this time using larger accretion disc profiles by setting $r_0 = 1.16 \times 10^{16}$cm (4.5 light-days) in equation (\ref{eq:parametrized}).
Larger microlensed sources result in a diminishing microlensing effect.
As shown in the right panel of Fig. \ref{fig:result_doubles_pdf}, the peak of each individual PDF now lies closer to $\mu_{\rm sad}/\mu_{\rm min} = 1.0$.
This may seem to have the same effect as increasing the smooth matter fraction, therefore introducing a degeneracy with the source size, however this is not the case: the height of the PDFs differs according to the smooth matter content in each mock population.
In fact, a similar trend is observed as before, i.e. for larger sources a lower smooth matter fraction leads to a lower peak of the PDF (located very close to $\mu_{\rm sad}/\mu_{\rm min} = 1.0$), in direct analogy to smaller sources having their peaks further away from $\mu_{\rm sad}/\mu_{\rm min} = 1.0$ as the smooth matter content decreases (the ordering of the PDFs in both panels of Fig. \ref{fig:result_doubles_pdf} is conserved).

\subsection{Measuring the IMF and the compact matter distribution}
As Euclid begins surveying the sky and discovering strong lenses, a single image of each system will be provided in each filter.
Based on these observations, one can calculate flux ratios for the multiply imaged quasars.
Will such flux ratios, especially the ones in the VIS filter, be enough to constrain the partitioning of matter in the lens and the IMF?

\begin{figure*}
	\includegraphics[width=\textwidth]{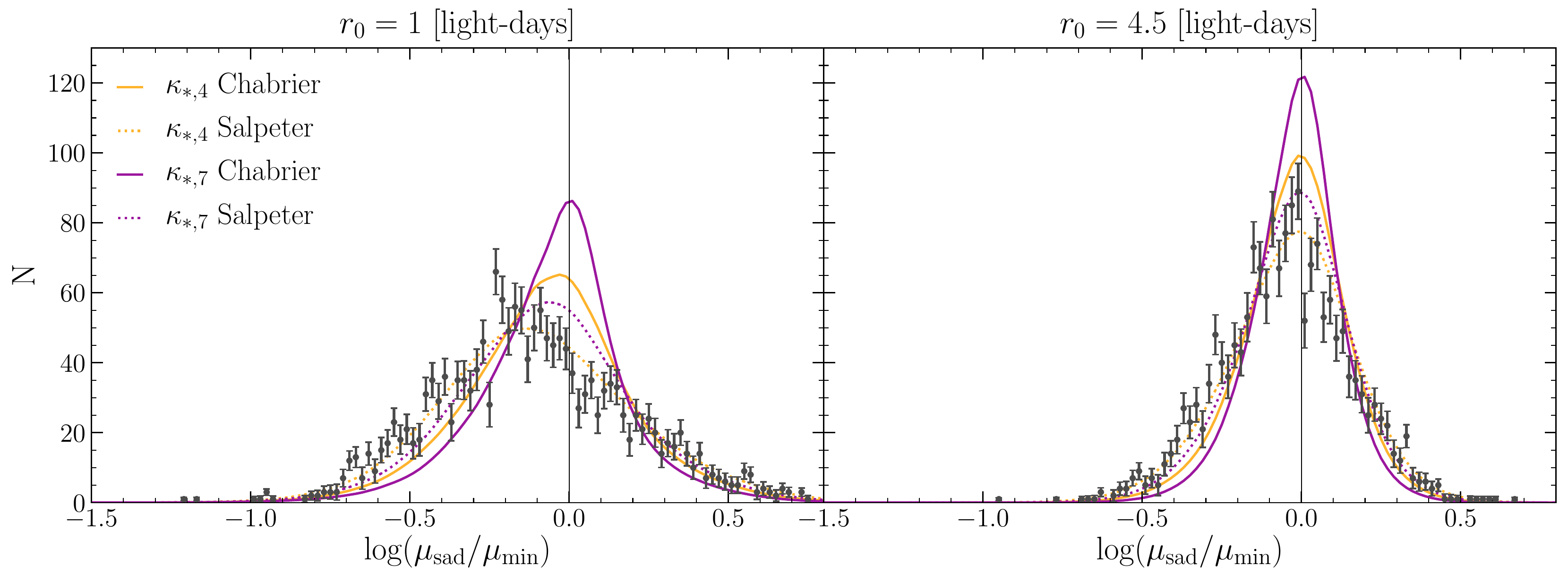}
	\caption{A set of mock flux ratio observations (points with errorbars) of 1774 CASTLES-like doubly imaged quasars at the wavelength of 900nm. A single flux ratio is assumed to be measured for each system (single-epoch observation) with a 10 per cent uncertainty. The mock lenses are assumed to have a Salpeter IMF and a Sersic profile with index 4 for the compact matter. The PDFs of the all the models (shown in Fig. \ref{fig:result_doubles_pdf}) are also shown after having been converted to number counts. Two populations of accretion discs are examined using equation (\ref{eq:parametrized}) with $r_0$ equal to 1 (left panel) and 4.5 (right panel) light-days. The vertical line indicates the value of the ratios in the absence of microlensing.}
	\label{fig:example_obs}
\end{figure*}

To answer this, mock sets of Euclid-like observations are created from simulated flux ratios.
The IMF is assumed to be either a Chabrier or a Salpeter one and the compact matter to have a Sersic profile with $n=4$ or $n=7$ \citep[typical for average-to-massive early-type galaxies; e.g.][]{Kormendy2016}, resulting in 4 different sets of mock observations.
The lenses are always assumed to follow the $\theta_{\rm Ein}/\theta_{\rm eff}$ relation from CASTLES.
The flux ratio PDF for each individual pair of images is sampled randomly once and the resulting frequency histogram of the observations is created.
An example of such a mock set of observations is shown in the Fig. \ref{fig:example_obs} for smaller ($r_0 = 1$ light-day in equation \ref{eq:parametrized}, left panel) and larger ($r_0 = 4.5$ light-days, right panel) accretion disc profiles.
Each mock set of observations is tested against all 4 possible combinations of IMFs and Sersic indices.
A $\chi^2$ statistic is used to measure the goodness-of-fit of each model to the mock data.
The goal is to test whether the model that was actually used to create the mock data is also the one that fits the data best (having the minimum $\chi^2$).

Four different collections of observed flux ratios are created, consisting of: the closest pairs of images in quadruple lenses with a fold configuration, the 4 pairs between a saddle-point and a minimum image in quadruple lenses of any configuration, and the double lenses with two different families of accretion disc profiles.
The number of measured flux ratios in each case is 188, 1044, and 1774, and the calculated $\chi^2$ values are shown in Table \ref{tab:chi2}.
These values are insensitive to the number of bins in the histogram of the observations or the precision of the flux ratio measurements.
The uncertainty of the measurement assigned to each individual flux ratio sets the uncertainty of the number counts in the histogram bins.
Each flux ratio was assumed to be a random variable with a fiducial standard deviation equal to 10 per cent.
Increasing or decreasing this uncertainty (e.g. in the reasonable range between 5 and 20 per cent), or even assuming a Poisson error for the histogram counts, has a minimal effect on the $\chi^2$ values.

The mock flux ratios obtained for doubly lensed quasars turn out to be those for which the underlying model is recovered successfully in every case, i.e the model with correct IMF and Sersic index assumptions is also the best-fitting one (having the minimum $\chi^2$ value), for both size parameters.
The model having a Salpeter IMF and a Serisc index $n=4$ is correctly recovered in all four sets of observations.
In the case of quadruple lenses, although the model with a Chabrier IMF and a Sersic index $n=7$ has the most different PDF compared to the others (see Fig. \ref{fig:result_realistic_all} and left panel of Fig. \ref{fig:result_realistic_closest}), one measurement of the flux ratio of each pair is not enough to uniquely recover it.
For example, the correct model is also the best-fitting one if 3 flux ratio measurements per image pair are used in the case of the closest image pairs, or 2 in the case of all the pairs between a saddle-point and a minimum image\footnote{The additional measurements per pair need to occur in different epochs and be uncorrelated with each other. This means that the observations have to be separated by enough time for the quasar accretion disc to cross a distance corresponding roughly to its own size. Such timescales can vary from months to years \citep{Mosquera2011b}, corresponding to sample sizes of 564 and 2088 flux ratios respectively.}.

\begin{table}
	\centering
	\caption{$\chi^2$ values obtained by comparing sets of mock Euclid data (rows) and different models (columns). There are 4 models having a different IMF and Sersic profile index $n$: Salpeter and $n=4$ (S4), Salpeter and $n=7$ (S7), Chabrier and $n=4$ (C4), Chabrier and $n=7$ (C7). The mock data are produced from one of these 4 models. Results are shown for four different sets of mock observations of image pairs, consisting of: the closest pair of images in quadruple lenses with a fold configuration, the 4 pairs between a saddle-point and a minimum image in quadruple lenses of any configuration, and the double lenses for sources with the parameter $r_0$ from equation (\ref{eq:parametrized}) set to 1 (small) and 4.5 (large) light-days respectively. If the minimum value of the $\chi^2$ (shown in bold) occurs on the diagonal of each sub-table then the best-fitting model is also the one used to create the mock data. This occurs only for the doubles for all the IMF, Serisc index, and source size combinations examined here.}
	\label{tab:chi2}
	\begin{tabular}{lcccc}
		
			 \multicolumn{5}{c}{Closest quads (188 flux ratios)}  \\
		data\textbackslash model & S4	& S7	& C4	& C7 \\
		\hline
		S4       & $\boldsymbol{79}$   & 84    & 91    & 107 \\
		S7       & $\boldsymbol{75}$   & 78    & 82    &  92 \\
		C4       & $\boldsymbol{69}$   & 70    & 74    &  82 \\
		C7       & $\boldsymbol{54}$   & $\boldsymbol{54}$    & 55    &  59 \\
		\hline
		
				 \multicolumn{5}{c}{Sad.-min. quads (1044 flux ratios)} \rule{0pt}{4ex} \\
		data\textbackslash model & S4	& S7 	& C4	& C7  \\
		\hline
		S4      & $\boldsymbol{47}$ 	& 84 	& 144 	& 349 \\
		S7 		& 51     & $\boldsymbol{44}$ 	&  73 	& 213 \\
		C4 		& 82     & $\boldsymbol{49}$ 	&  55 	& 144 \\
		C7 		& 154    & 79 &  $\boldsymbol{44}$ 		&  56 \\
		\hline
						
				 \multicolumn{5}{c}{Doubles small (1744 flux ratios)} 	\rule{0pt}{4ex}    	\\
		data\textbackslash model & S4	& S7	& C4	& C7 	\\
		\hline
        S4		& $\boldsymbol{126}$	& 229	& 377	& 856	\\
        S7		& 161   & $\boldsymbol{137}$ 	& 196	& 483	\\
        C4		& 229   & 115	& $\boldsymbol{110}$	& 273	\\
        C7		& 613   & 307	& 180	& $\boldsymbol{129}$	\\
		\hline
		
				\multicolumn{5}{c}{Doubles large (1744 flux ratios)} 	\rule{0pt}{4ex}    	\\
		data\textbackslash model & S4	& S7	& C4	& C7 	\\
		\hline
		S4		& $\boldsymbol{96}$	& 155		& 254	& 516	\\
		S7		& 131   & $\boldsymbol{109}$ 	& 153	& 324	\\
		C4		& 221   & 106	& $\boldsymbol{97}$		& 192	\\
		C7		& 616   & 295	& 164	& $\boldsymbol{99}$		\\
		\hline
						 
	\end{tabular}
\end{table}

\section{Discussion and conclusions}
\label{sec:discuss}
In this work, the theoretical properties of microlensing flux ratios of large collections of multiply imaged quasars are explored.
Mock observations are used, adjusted to the Euclid shallow survey specifications in the VIS filter.
The flux ratio distributions are analyzed and found to depend critically on the value of the smooth matter fraction $s$, which in turn depends on the assumptions on the IMF and compact matter distribution.
The differences between such distributions are measurable and hence can provide constraints on the IMF and mass components of galaxies out to the redshifts of the lenses.

A mock population of lenses expected for the Euclid survey is extracted from the \citet{Oguri2010} catalog.
The main characteristic of these lenses is their isothermal total mass profile that constrains the values of $\kappa$ and $\gamma$ at the location of the multiple images on a roughly straight line (see Fig. \ref{fig:pspace}); it can be easily shown that $\kappa= | \gamma | $ for a pure SIE (see equations \ref{eq:gamma-sie-components}).
For the $\kappa,\gamma$ combinations explored here, it is determined that the value of $s$ plays the most important role.
In fact, the $\kappa,\gamma$ of the mock lenses are marginalized over in the analysis.
Nevertheless, more advanced mock samples with varying lens potential slopes \citep[e.g.][]{Keeton2001b} would be worth investigating in the future, e.g. in order to examine the impact of slope evolution with redshift.

The light profile of the mock lenses is constrained by two different datasets of observed lenses, i.e. the CASTLES sample, consisting purely of lensed quasars, and the SLACS sample, containing massive galaxy-galaxy lenses.
The ratio of the effective radius of the light profile to the Einstein radius of the lens, $\theta_{\rm Ein}/\theta_{\rm eff}$, which is a measure of the extent of the light profile over the total mass profile, is quite different in the two samples.
This ratio is affected by the measurements of the source and lens redshifts and the velocity dispersion of the lens that go into calculating $\theta_{\rm Ein}$ (see equation \ref{eq:theta_rein}), and the value of $\theta_{\rm eff}$.
The two samples have quite different distributions of the $\theta_{\rm Ein}/\theta_{\rm eff}$ ratio (see histograms in Figs. \ref{fig:reff_rein} and \ref{fig:ratio_rein_reff}), with SLACS having more extended light profiles with respect to the Einstein radius.
This leads to more mass in the form of compact objects at the locations of the multiple quasar images and therefore a lower value for $s$ with respect to a CASTLES-like sample.
Different combinations of IMF and compact matter profiles are almost indistinguishable when the $\theta_{\rm Ein}/\theta_{\rm eff}$ relation from SLACS is used (see Fig. \ref{fig:result_realistic_closest}).
CASTLES-like lenses are found to be most promising for measuring $s$ and constraining the IMF and compact matter profiles.

The microlensing flux ratio distributions of samples of hundreds of lensed quasars are found to vary continuously as a function of the smooth matter fraction $s$ between two extremes.
On one hand, the distribution has a broad peak at roughly 0.3 times the macro-magnification ratio for collections of image pairs with $s \lesssim 0.95$.
On the other hand, the distribution is sharply peaked at the macro-magnification ratio (no microlensing) for $s \gtrsim 0.95$.
A smooth transition from the former to the latter occurs as a function of the number of pairs with at least one of the images having $s \ge 0.95$.
This is demonstrated in Fig. \ref{fig:result_constants_semirandom}, for accretion discs with sizes determined from equation (\ref{eq:parametrized}) with $r_0 = 1$ light-day.
Larger accretion discs having $r_0 = 4.5$ light-days are examined for doubly imaged quasar populations.
In this case, the flux ratio distributions are all peaked around the macro-magnification ratio but their height varies (see right panel of Fig. \ref{fig:result_doubles_pdf}).
The same trend is observed as before, this time regarding the height of the distributions which varies smoothly as a function of the number of systems with at least one of the images having $s \ge 0.95$.

The value of $s$ at the locations of the multiple images depends on the partition of matter in the lens in smooth and compact components.
More concentrated compact matter profiles produce a higher $s$ for the lensed images.
This is explored using Sersic profiles for the compact matter component, where a higher Sersic index ($n=7$) always leads to higher $s$ than a lower one ($n=4$, e.g. see Figs. \ref{fig:result_realistic_all} and \ref{fig:result_doubles_pdf}).
Additionally, a Salpeter (i.e. bottom-heavier) IMF implies more lower mass lenses than a Chabrier one, increasing the number of compact objects and hence lowering the value of $s$.
Therefore, the case with a Salpeter IMF and a Sersic index of $n=4$ is the closest to the one extreme of the flux ratio distributions, i.e. it shows a broad peak around 0.3 times the macro-magnification ratio, while the case with a Chabrier IMF and a Sersic index of $n=7$ lies at the other extreme, i.e. having a sharp peak at the value of the macro-magnification ratio.
The cases of a Salpeter IMF with an index of $n=7$ and a Chabrier IMF with an index of $n=4$ are neither dominated by nor devoid of pairs with at least one image having $s \ge 0.95$.
Hence, the corresponding flux ratio distributions are found somewhere in between and are harder to distinguish from each other (see Table \ref{tab:chi2}).

It is found that the best systems to constrain the smooth matter fraction of the lenses are the doubly imaged quasars with CASTLE-like lenses.
This is largely due to their larger numbers with respect to quadruple lenses, but also due to the generally wider separation of the images, reflected in the more extended range of their $\kappa,\gamma$ (see Fig. \ref{fig:pspace}).
The closest image pairs in fold quadruple lenses (a factor of ten fewer than the doubles) perform the worst in this kind of experiment, despite their weaker dependence on the macromodel of the lens.

Flux ratio observations can be contaminated by other factors mimicking microlensing, e.g. uncertainties in the macromodel, substructure, and/or the state of the variable background quasar.
Indeed, the larger time delays associated with doubles increase the chance of the quasar being seen in a different state in the two images, unless it is already in a quiescent state anyway.
Nevertheless, the effect of the first two factors could be mitigated by using the Euclid infrared filters\footnote{In this case, the observational challenge would be to deblend close separation images in the infrared instruments, which have a 0.3 arcsec resolution as opposed to the 0.1 arcsec of the VIS instrument.} or infrared spectrometer \citep[see][for an example of using slitless spectroscopy to derive flux ratios]{Nierenberg2017}.
The source in such wavelengths is large enough for any microlensing effect to be negligible, therefore, the un-microlensed flux ratios could be determined without having to produce an accurate macromodel or assuming substructure (this is similar to the technique employed by \citealt{Moustakas2003} and \citealt{Mediavilla2009} using narrow emission lines).
Based on such baseline flux ratios the microlensing deviations in the VIS filter could be measured.
Finally, if multi-epoch observations are available, e.g. from follow-up campaigns, providing larger numbers of independent flux ratios per system, the evolution of smooth matter fractions as a function of redshift could be constrained.

In this work, mock lens quasar populations are assumed having single values for the Sersic index $n$ of the compact matter profile and the accretion disc size parameter $r_0$.
A more realistic case would be considering population mixtures.
Together with contamination of flux ratios due to non-microlensing factors (and to a lesser degree the SIE lens potential of the lenses, see above, and the wide range of the Euclid VIS filter, see Section \ref{sec:results}), these are the two main limitations of this work.
The conclusions presented in this section should therefore be viewed taking these assumptions into account.

The approach adopted in this work is to obtain key properties of lensing galaxies in a `quick-and-dirty' way, and can be directly applied to the data products of the Euclid survey (no follow-up of the systems).
There is no need for any modelling of the lenses, measuring the redshifts, or fitting of the lens light profiles (although any such corollary information would certainly be beneficial).
The only data required is a measurement of the flux ratios at the locations of the multiple images, which may or may not be contaminated by other factors.
In any case, simply due to the large number of systems (1774 doubles examined here) such effects may wash out.
This approach may be promising for other studies of large collections of lenses, e.g. modelling of galaxy-galaxy lenses with or without substructure, using flux ratios to determine accretion disc properties, or investigating microlensing light curves.

\section*{Acknowledgements}
The author would like to thank M. Oguri for his help in using the \citet{Oguri2010} catalog of mock lenses, C. Spingola for fitting lens models to multiply imaged quasars, and L. V. E. Koopmans for helpful discussions during the various stages of this work.
Thanks goes to P. Bonfini for reading draft versions of the present manuscript and providing comments that improved the quality of the final result.
The author is supported through an NWO-VICI grant (project number 639.043.308).
This work was performed on the gSTAR national facility at Swinburne University of Technology.
gSTAR is funded by Swinburne and the Australian Government's Education Investment Fund.

\bibliographystyle{mnras}
\bibliography{ml_surveys}

\appendix
\section{Additional tables and figures}

\begin{table}
	\centering
	\caption{Parameters of the fiducial quadruply imaged lens system of Fig. \ref{fig:example}. These are: the velocity dispersion, $\upsilon_{\rm disp}$, the dynamical normalization factor, $\lambda$, the position angle of the lens ellipsoid, $\phi_{\rm L}$ (counter-clockwise from the x-axis), the axis ratio of the SIE mass model, $q$, and the magnitude and direction of the external shear, $\gamma^{\rm ext}$ and $\phi_{\rm \gamma}$. $x,y$ are the lens center and point-source coordinates, and $z$ their redshifts.}
	\label{tab:example}
	\begin{tabular}{ccc}
		parameter				& value		& units		\\
		\hline
		$\upsilon_{\rm disp}$ 	& 204		& km/s		\\
		$\lambda$				& 1.002		&			\\
		$\phi_{\rm L}$			& 19 		& deg		\\
		$q$						& 0.698		&			\\
		$\gamma_{ext}$			& 0.042		&			\\
		$\phi_{\rm \gamma}$		& -118	 	& deg		\\
		$x_{\rm L}$				& 0.0		& arcsec	\\
		$y_{\rm L}$				& 0.0	 	& arcsec	\\
		$z_{\rm L}$				& 0.58		&			\\
		$x_{\rm S}$				&  0.081	& arcsec	\\
		$y_{\rm S}$				& -0.018 	& arcsec	\\
		$z_{\rm S}$				& 2.31		&			\\
		\hline
	\end{tabular}
\end{table}

\begin{table}
	\centering
	\caption{Properties of the multiple images of the fiducial system of Fig. \ref{fig:example}. Coordinates $x$ and $y$ are in arcsec.}
	\label{tab:example_images}
	\begin{tabular}{lccccc}
		image			& $x$ 		& $y$  		& $\kappa$	& $\gamma$	& $\mu$		\\
		\hline
		A (minimum)		&  0.413	& -0.718 	& 0.38		& 0.40		&  4.46		\\
		B (saddle-point)& -0.615	& -0.050 	& 0.70		& 0.66		& -2.89		\\
		C (minimum)		&  0.311	&  0.738 	& 0.44		& 0.46		&  9.80		\\
		D (saddle-point)&  0.708	&  0.317 	& 0.57		& 0.53		& -10.42	\\		
		\hline
	\end{tabular}
\end{table}

\begin{figure}
	\includegraphics[width=0.47\textwidth]{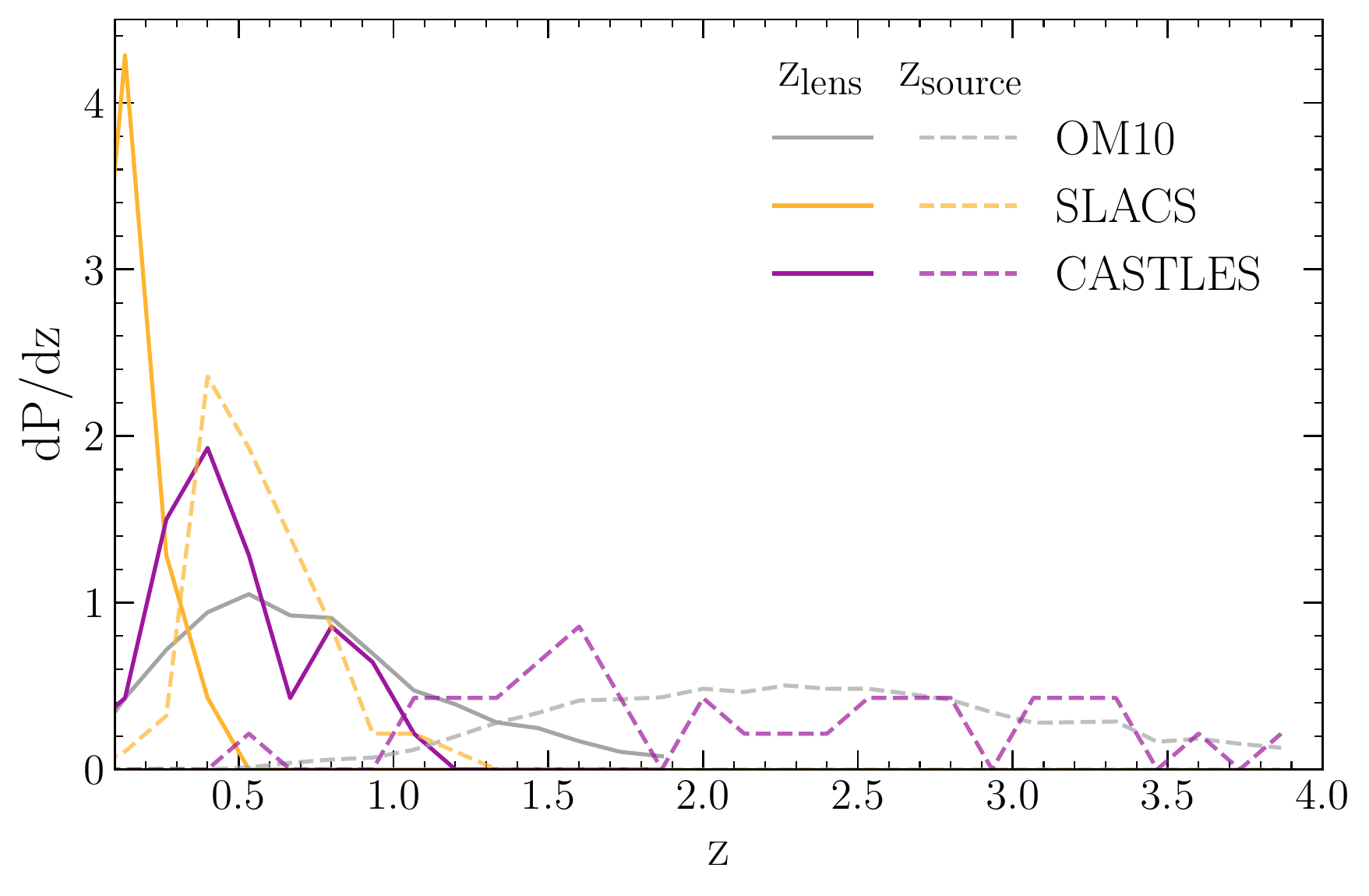}
	\caption{Probability density distributions for the source (dashed line) and lens (solid line) redshifts from the SLACS and CASTLES samples of observed lenses, and the OM10 sample of mock lenses used here.}
	\label{fig:z_hist}
\end{figure}

\begin{figure}
	\includegraphics[width=0.47\textwidth]{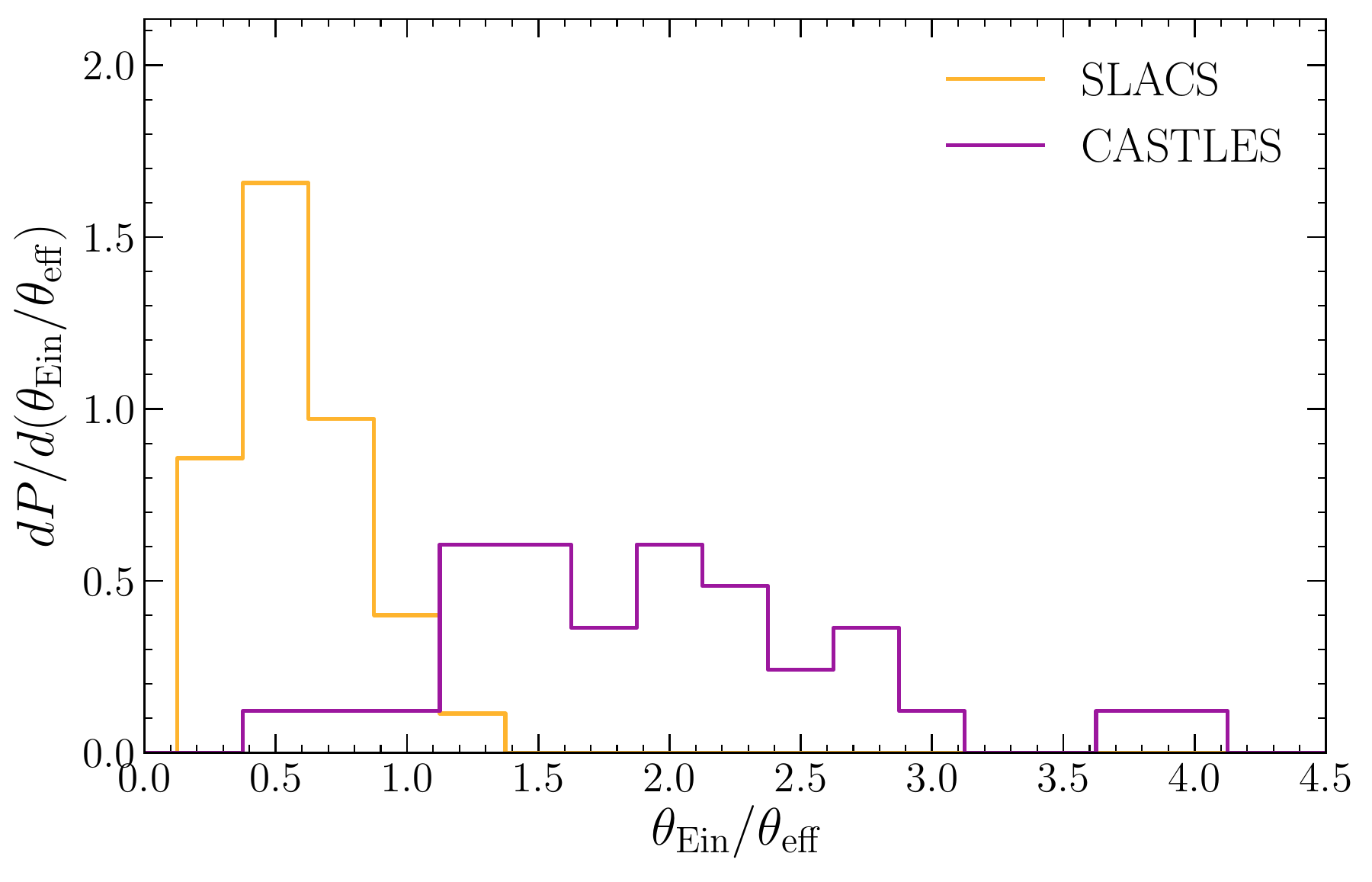}
	\caption{Probability density distributions for the $\theta_{\rm Ein}/\theta_{\rm eff}$ ratio for the SLACS and CASTLES samples of observed lenses. The individual $\theta_{\rm Ein}$ and $\theta_{\rm eff}$ value distributions are shown in the histograms of Fig. \ref{fig:reff_rein}.}
	\label{fig:ratio_rein_reff}
\end{figure}

\end{document}